\documentclass[a4paper,11pt]{article}
\usepackage{jcappub} % for details on the use of the package, please see the JINST-author-manual
\usepackage{lineno}
\usepackage{amssymb}
\usepackage{makecell}
\usepackage{hyperref}
\usepackage{float}
\usepackage{booktabs}
\usepackage{amsmath}
\usepackage{comment}
\usepackage{makecell}

\newcommand{\lb}{\textit{LiteBIRD}}

\newcommand{\sm}[1]{{#1}}

\title{Combining Systematic Effects in CMB Polarization Experiments through map-based simulations: application to {\mdseries\itshape LiteBIRD}'s HWP non-idealities and detectors non-linearity}

\author[1, 2, 3]{Silvia Micheli,}
\author[1,2]{Francesco Piacentini,}
\author[4,5]{Tommaso Ghigna,}
\author[6,7]{Alessandro Carones,}
\author[8]{Alessandro Novelli,}
\author[1,2]{Giampaolo Pisano,}
\author[1,2]{Giulia Barbieri Ripamonti,}
\author[1,2]{Alessandro Coppolecchia,}
\author[1,2]{Paolo de Bernardis,}
\author[1,2]{Silvia Masi,}
\author[1,2]{Andrea Occhiuzzi,}
\author[1,2]{Alessandro Paiella,}
\author[1,2]{Simone Stellati}

\affiliation[1]{Dipartimento di Fisica, Univ. La Sapienza, P.le A. Moro 2, Roma, Italy}
\affiliation[2]{INFN Sezione di Roma, P.le A. Moro 2, I-00185 Roma, Italy}
\affiliation[3]{Université Grenoble Alpes, CNRS, LPSC-IN2P3, 53, Avenue des Martyrs, 38000 Grenoble, France}
\affiliation[4]{International Center for Quantum-field Measurement Systems for Studies of the Universe and Particles (QUP), High Energy Accelerator Research Organization (KEK), Tsukuba, Ibaraki 305-0801, Japan}
\affiliation[5]{Kavli Institute for the Physics and Mathematics of the Universe (Kavli IPMU, WPI), UTIAS, The University of Tokyo, Kashiwa, Chiba 277-8583, Japan}
\affiliation[6]{International School for Advanced Studies (SISSA), Via Bonomea 265, Trieste 34136, Italy}
\affiliation[7]{INFN Sezione di Trieste, 
Via Valerio 2, Trieste 34127, Italy}
\affiliation[8]{Dipartimento di Fisica, Univ. of Milano-Bicocca, Piazza della Scienza 3, 20126, Milano,
Italy}

\emailAdd{silvia.micheli@lpsc.in2p3.fr}

\abstract{
We quantify the impact of coupled instrumental systematics on next-generation CMB polarization experiments targeting primordial $B$-mode polarization, with a focus on the forthcoming \textit{LiteBIRD} satellite mission. We study the interplay between the non-linear response of Transition-Edge Sensor (TES) bolometers and Half-Wave Plate (HWP) non-idealities, in particular synchronous signals arising from differential emissivity. We develop a map-based formalism that captures the resulting intensity-to-polarization leakage by explicitly solving the binning map-making equations, avoiding the need for computationally expensive time-ordered data simulations. We apply this framework to \textit{LiteBIRD}, adopting its baseline scanning strategy and frequency configuration, and perform analyses at both single- and multi-frequency levels, including Galactic foregrounds and blind component separation. We find that while detector non-linearity and HWP non-idealities individually induce negligible bias on the tensor-to-scalar ratio $r$, their coupling can generate non-trivial contamination, driven by large signals such as the solar dipole and amplified in high-frequency channels by foreground leakage. From these results, we derive joint requirements on detector non-linearity and HWP differential emission, and discuss their implications for the instrument design and calibration strategy of \textit{LiteBIRD} and future CMB polarization missions targeting $r \lesssim 10^{-3}$.
}

\begin{document}
\maketitle
\flushbottom

%For internal references use label-refs: see section~\ref{sec:intro}.
%Bibliographic citations can be done with "cite": refs.~\cite{a,b,c}.

\section{Introduction}
\label{sec:intro}
The observation of the Cosmic Microwave Background (CMB) is one of the most powerful probes for understanding the early Universe, providing key insights into the $\Lambda$CDM model through missions such as COBE~\cite{Smoot:1992}, WMAP~\cite{Hinshaw:2013}, and Planck~\cite{PlanckCollaboration:2020I}. Temperature anisotropies in the CMB have played a crucial role in shaping our cosmological understanding, while polarization anisotropies encode additional, complementary information that remains only partially explored. In particular, the measurement of primordial $B$-mode polarization offers a unique opportunity to probe cosmic inflation, a period of rapid expansion in the early Universe that sourced initial conditions for cosmological perturbations via primordial quantum fluctuations. Inflationary models predict a stochastic background of gravitational waves, which would leave a distinctive $B$-mode signature in the CMB polarization~\cite{Seljak:1997}. The amplitude of these tensor perturbations is parameterized by the tensor-to-scalar ratio, \(r\), whose measurement would place strong constraints on inflationary physics~\cite{Kamionkowski:2016}. \\ Detecting primordial $B$-modes remains a major challenge due to their elusiveness, requiring next-generation CMB experiments with unprecedented sensitivity. Ground-based observatories such as the Simons Observatory~\cite{Ade:2019}, alongside space missions like \lb\ \cite{LiteBIRDCollaboration:2023} and PICO \cite{Hanany:2019}, aim to push the limits of detection, either revealing primordial $B$-modes or further tightening constraints on \(r\). While current observations have only set upper bounds, with the tightest being \(r < 0.032\) (95\% CL)~\cite{Tristram:2022}, future surveys will provide a decisive test of inflationary models by probing the $B$-mode power spectrum across all scales, from the large angular scales, where primordial B-modes could dominate over lensing-induced one, to smaller scales where delensing techniques are used to mitigate the lensing background \cite{Smith:2012}. To reach this remarkable goal, modern CMB experiments have to face several challenges, including dealing with Galactic foregrounds contamination and control over systematic effects. In this work, we focus on the future space experiment \lb\ (Lite (Light) satellite for the studies of $B$-mode polarization and Inflation from cosmic microwave background Radiation Detection)~\cite{LiteBIRDCollaboration:2023, Ghigna:2024}, which has been selected by the Japanese Aerospace Exploration Agency (JAXA) to observe the full-sky for 3 years, aiming to reach a final sensitivity of $\delta r < 10^{-3}$ by measuring both the recombination and the reionization bumps. To achieve this goal, \lb\ will observe the sky over a wide frequency range, from 34 to 448 GHz employing a Low-, a Medium- and a High- Frequency Telescope (LFT, MFT and HFT, respectively), to characterize the foreground emission and increase the sensitivity in the CMB channels. A common choice for next-generation CMB experiments, including \lb, is to employ a continuously rotating Half-Wave Plate (HWP) to modulate the incoming polarization, mitigate the effect of $1/f$
noise on the polarization signal, and minimize differential systematics \sm{arising from pair-differencing between orthogonal detectors~\cite{Wallis:2017}}. However, non-idealities of the HWP can introduce systematic effects that must be carefully addressed. Several works~\sm{\cite{Stellati:2025, Monelli:2023, Ritacco:2017, Salatino:2018, Takakura:2019}} have been conducted to assess the impact of HWP systematics on the reconstruction of CMB maps. \sm{\lb\ will be equipped with highly sensitive Transition-Edge Sensor (TES) bolometers to measure the polarized signal coming from the sky.} In this paper, we study the interplay between spurious signals generated by HWP non-idealities and the non-linear response of TES detectors. In a previous work~\cite{Micheli:2024}, a dedicated pipeline based on full time-ordered data (TOD) simulations was developed to set requirements on the knowledge of TES non-linearity for the \lb\ mission. Here, we extend that framework by introducing a faster, map-based formalism that captures the same leakage effects without going through computationally expensive TOD simulations. Following the approach of~\cite{ThuongHoang:2017,Banerji:2019,McCallum:2022}, we construct templates that allow us to reproduce the effect of detector non-linearity directly at the map level, by explicitly solving the binning map-making equation. This method enables efficient simulation of leakage from TES non-linearity combined with HWP imperfections, while maintaining consistency with TOD-based results. In this work, we adopt the baseline \lb\ scanning strategy and design as presented in~\cite{LiteBIRDCollaboration:2023} to validate our method and derive instrumental requirements. The \lb\ baseline design is undergoing a redesign phase to consolidate the mission’s feasibility while maintaining its primary scientific goals. Nevertheless, the approach presented here is general and applicable to any satellite configuration. For this reason, we retain the baseline design as a reference, allowing for a direct comparison with previous work~\cite{Micheli:2024}. 
Previous CMB experiments such as POLARBEAR and EBEX~\cite{Takakura:2017,Didier:2019} have shown that HWP non-idealities can generate strong spurious signals that \sm{can further contaminate the data due to the interplay with TES non-linearity}. Although \textit{LiteBIRD} \sm{design is different}, similar effects are expected~\cite{Micheli:2024}, making it essential to quantify their impact. 
The paper is organized as follows. In section.~\ref{sec:formalism} we present the formalism for building template maps that account for detector non-linearity, and validate it against full TOD simulations. In section.~\ref{sec:simulations} we describe the data reduction pipeline, showing results for different combinations of systematics' amplitudes and their impact on the recovery of the tensor-to-scalar ratio, including the role of component separation. Finally, section.~\ref{sec:conclusions} summarizes our findings and provides design recommendations to mitigate these systematic effects in the specific case of \textit{LiteBIRD}.

\section{Formalism}\label{sec:formalism}
\paragraph{Detector response and non-linearity} The \textit{LiteBIRD} mission utilizes a large number of Transition-Edge Sensors (TES) for precise CMB polarization measurements. A TES operates by measuring the change in current, $I$, induced by a change in resistance, \sm{whose value} is highly sensitive to absorbed optical power, $P_\text{opt}$. This relationship is inherently non-linear \cite{Irwin:2005, Ghigna:2023, deHaan:2024}. %The responsivity, $S \equiv \partial I / \partial P_{\text{opt}}$, depends on the absorbed optical power.
We model this dependence with a second-order expansion of the gain:
\begin{equation}\label{eq:nonlin-tod}
d_{\text{NL}}(t) = d(t)\,\big(1 + g_1\, d(t)\big),
\end{equation}
where $d(t)$ is the measured signal and $g_1$ is the non-linearity parameter, typically negative. This quadratic form is the standard model for characterizing the non-linear response of TES detectors and has been extensively used in the literature. For instance, studies for POLARBEAR \cite{Takakura:2017} and Simons Observatory \cite{Salatino:2018} adopt this notation, reporting typical values around $g_1 \approx -4 \times 10^{-3} \mathrm{K}^{-1}$. We assume here the linear gain to be perfectly calibrated, i.e., equal to one. Possible time-dependent features of the detectors are usually encoded in a non-linear time constant term \cite{Takakura:2017b}. In our case, we do not include this kind of effect in order to apply the map-based formalism, which does not capture drift effects, and to isolate the effect of a non-linear gain coupled with spurious signals.
\newline
For an ideal, rotating Half-Wave Plate (HWP), the measured signal for a single detector, $j$, is a function of the incoming Stokes parameters $I$, $Q$, and $U$:
\begin{equation}\label{eq:tod}
d_j = I(p) + Q(p) \cos(2 \chi_j) + U(p) \sin(2 \chi_j) + n_j
\end{equation}
Here, $ 2\chi = 4\theta - 2\psi $, where $\theta = \omega_{\mathrm{HWP}} t$ and $\psi$ represent respectively the orientation angles of the HWP and detector relative to the sky reference frame, whose $z$-axis is aligned with the telescope boresight and $x$-axis defines the reference direction. The $n_j$ term indicates the noise contribution to the signal, considered white in this analysis. Here, we consider that the detectors in the same frequency channel share the same noise level. While we include the noise in the full map-making process, we omit it from the intermediate analytical expansions of the non-linear terms. This is because the higher-order noise contributions, specifically the noise-squared term ($n^2$) and the cross-terms between noise and signal, are several orders of magnitude smaller than the signal-squared terms and do not significantly impact the bias estimation. This simplification allows for a clearer derivation of the templates without loss of accuracy in the final results. We can now explicitly write eq \ref{eq:nonlin-tod} to see that non-linearity introduces a signal variation
\begin{equation}
    \delta d_{\text{NL}} = g_1 (I+Q\cos{2\chi}+U\sin{2\chi})^2
\end{equation}
which results in leakage from I to Q and U maps. The estimated Stokes parameters ($\hat{I}$, $\hat{Q}$, $\hat{U}$) for each sky pixel $p$ are derived using a binning map-making approach \cite{Tegmark:1997}:
\begin{equation}\label{eq:mapmaking}
\hat{S}=\begin{pmatrix}
\hat{I} \\
\hat{Q} \\
\hat{U}
\end{pmatrix}
= M^{-1}
\begin{pmatrix}
\langle d_j \rangle \\
\langle d_j \cos (2\chi_j) \rangle \\
\langle d_j \sin (2\chi_j) \rangle
\end{pmatrix}
\end{equation}
where $M$ is the cross-linking matrix:
\begin{equation}
M = \begin{pmatrix}
1 & \langle\cos (2\chi_j) \rangle &\langle\sin (2\chi_j) \rangle \\
\langle\cos (2\chi_j) \rangle & \langle\cos^2 (2\chi_j) \rangle  & \langle\cos (2\chi_j)\sin(2\chi_j) \rangle\\
\langle\sin (2\chi_j) \rangle & \langle\sin (2\chi_j)\cos (2\chi_j) \rangle & \langle\sin^2 (2\chi_j) \rangle \, .
\end{pmatrix}
\end{equation}
The angle brackets $\langle \rangle$ indicate the average over the hits on the same sky pixel, and $\hat I$, $\hat Q$, $\hat U$ are the estimated Stokes parameters. In general, we can simplify the expression for the $M$ matrix if we consider pairs of perfectly orthogonal detectors and an optimal scanning strategy. In this context, an optimal strategy is defined by a coverage that ensures a uniform distribution of polarization crossing angles for each observed pixel\footnote{Mathematically, this corresponds to a configuration where the orientations of the polarimeters are evenly distributed over $180^\circ$, satisfying $\alpha_p = \alpha_1 + (p - 1) \frac{\pi}{n}$ with $n \geq 3$~\cite{Couchot:1999}.}. This is the case for most upcoming CMB experiments. According to~\cite{Couchot:1999}, with this optimal arrangement $M^{-1}$ reduces to $diag(1, \, 1/2, \, 1/2)$. However, since we want to know which new terms arise from adding non-linearity, we keep a general form for the cross-link matrix. The full expressions are listed in appendix \ref{sec:appendix}. 
%The estimated Stokes parameters, including this non-linear contamination, can be expressed as \sm{(see appendix \ref{sec:appendix} for the full derivation)}:
%\begin{equation}\label{eq:contStokes-inthetext}
%\hat{S} = \frac{1}{\det(M)}
%\begin{pmatrix}
%(CE-D^2) \langle d_{NL} \rangle + (BD-AE) \langle d_{NL}\cos{2\chi} \rangle + (AD-BC) \langle d_{NL}\sin{2\chi} \rangle \\
%(BD-AE) \langle d_{NL} \rangle + (E-B^2) \langle d_{NL}\cos{2\chi} \rangle + (AB-D) \langle d_{NL}\sin{2\chi} \rangle \\
%(AD-BC) \langle d_{NL} \rangle + (AB-D) \langle d_{NL}\cos{2\chi} \rangle + (C-A^2) \langle d_{NL}\sin{2\chi} \rangle
%\end{pmatrix}
%\end{equation}
\paragraph{\sm{Realistic} HWP} A realistic HWP can produce strong parasitic signals, known as HWP synchronous signals (HWPSS), at harmonics of its rotation frequency, $f_{\text{HWP}}$. These can arise from intrinsic HWP differential optical properties and imperfections in the anti-reflection coating or non-normal incidence angles. At non-zero temperature, an HWP inevitably emits some unpolarized thermal radiation with intensity as a grey-body, according to:
\begin{equation}
    \varepsilon(\nu) B_\textsc{hwp}(\nu, T_\textsc{hwp}) = \varepsilon(\nu) \frac{2h\nu^3}{c^2} \frac{1}{e^{\frac{h\nu}{k_BT_\textsc{hwp}}}-1}\,,
\end{equation}
where $T_\textsc{hwp}$ is the HWP physical temperature and $\varepsilon(\nu)$ is the (dimensionless) total emissivity, which is generally frequency-dependent. The optical properties can also differ between the ordinary and extraordinary axes. For this reason, the HWP will also exhibit some differential emissivity, meaning it will emit slightly more radiation polarized along one axis than the other. We refer to this quantity as $\Delta \varepsilon$ throughout this paper. In the HWP’s own reference frame, with the $x$ and $y$ axes aligned to the ordinary and extraordinary axes respectively, the emitted radiation can be represented as the Stokes vector 
\begin{equation}
    \mathbf{S}_\textsc{hwp}(\nu) = I_\textsc{hwp}(\nu, T_\textsc{hwp})\begin{pmatrix}
        \varepsilon(\nu) \\ \Delta\varepsilon(\nu) \\ 0
    \end{pmatrix}\,,
\end{equation}
where $\Delta\varepsilon(\nu)$ parametrizes the polarized component of the emission, and $I_\textsc{hwp}$ denotes the blackbody intensity corresponding to the HWP temperature, expressed in CMB temperature units. \\ To simplify the model, we consider band-averaged quantities. So, for the $k$-th channel, we will express the HWP emissivity along each optical axis as the weighted mean:
\begin{equation}
    \varepsilon_k =\frac{\int_{\nu_i}^{\nu_f} \varepsilon(\nu) \frac{2h\nu^3}{c^2} \frac{1}{e^{\frac{h\nu}{k_B T_\textsc{hwp}}}-1}}{\int_{\nu_i}^{\nu_f} \frac{2h\nu^3}{c^2} \frac{1}{e^{\frac{h\nu}{k_B T_\textsc{hwp}}}-1}}\,,
\end{equation}
where $\nu_i$ and $\nu_f$ refer to the lowest and the highest frequencies of the $k$-th channel, respectively.
Analogous definitions hold for $\Delta \varepsilon_k$ and $I_{\textsc{hwp},k}(T_\textsc{hwp})$. These can be computed through a sensitivity calculation procedure \cite{Hasebe:2023} and converted to temperature units for TOD simulations as described in section 3.1 of \cite{Micheli:2024}. Thus, the HWP contribution to the TOD for channel $k$ can be written as:
\begin{equation}\label{eq:d_hwp_k}
d_{\textsc{hwp},k} = \varepsilon_k I_{\textsc{hwp},k}(T_\textsc{hwp}) + \Delta \varepsilon_k I_{\textsc{hwp},k}(T_\textsc{hwp}) \cos{\Theta}\,,
\end{equation}
Where $\Theta = 2(\theta-\xi_{\mathrm{det}})$ is the HWP rotation angle in the instrument frame, $\xi_{\mathrm{det}}$ being the detector polarization angle\footnote{We recall that the additional rotation by the angle $\xi_{\mathrm{det}}$ is applied to determine the contribution of the HWP differential emission, which lives in the HWP's own reference frame, to the TOD.}. The first term is unpolarized thermal emission, the second is polarized, modulated at twice the HWP-detector angle. In this work, we focus on the impact of the second term. Instead, the role of differential transmission and its interplay with the sky signal are discussed in \cite{Patanchon:2024} and will be addressed in future work. We focus here on differential emission because, as shown in \cite{Micheli:2024}, it can produce significantly larger contamination when coupled with detector non-linearity. \sm{Therefore, by combining eqs.~\eqref{eq:tod} and \eqref{eq:d_hwp_k}, we can include in the TOD an additional term originating from HWP differential emission:}
\begin{equation}\label{eq:deltad2f}
\delta d^{2f} = A_{2f} \cos{\Theta}\,,
\end{equation}
where $A_{2f}$ is the HWPSS amplitude. The units of $A_{2f}$ depend on the chosen units for the TOD; in this paper, we use $\mathrm{K_{CMB}}$ units. As mentioned, the expected values of $A_{2f}$ can be calculated once the HWP model is provided. %In this work, we adopt some typical values for $A_{2f}$, based on the LiteBIRD design as described in~\citep{Takaku:2023, Giardiello:2022} and on the signal amplitudes derived in~\citep{Micheli:2024}. 
%Figure \ref{fig:hwp-trans} shows the different values of transmittance along the two optical axes for realistic HWPs. Note that the different behavior between the LFT and MHFT plates is due to their different nature. In fact, LFT employs a sapphire HWP, whereas MHFT uses a metal-mesh HWP \cite{Pisano:2022, Takaku:2023}. 

%\begin{figure}
%    \centering
%    \includegraphics[width=0.8\linewidth]{new_figs/Txy-allTelescopes.png}
%    \caption{Transmittance of the five-layer sapphire HWP for LFT (coral) and the metal-mesh HWP for MFT and HFT (red and blue, respectively) as a function of frequency. Solid and dashed lines indicate the two optical axes. Further details can be found in %\cite{Monelli:2024, Pisano:2022}.}
%    \label{fig:hwp-trans}
%\end{figure} 

\begin{figure}[]
\centering
\includegraphics[width=0.8\textwidth]{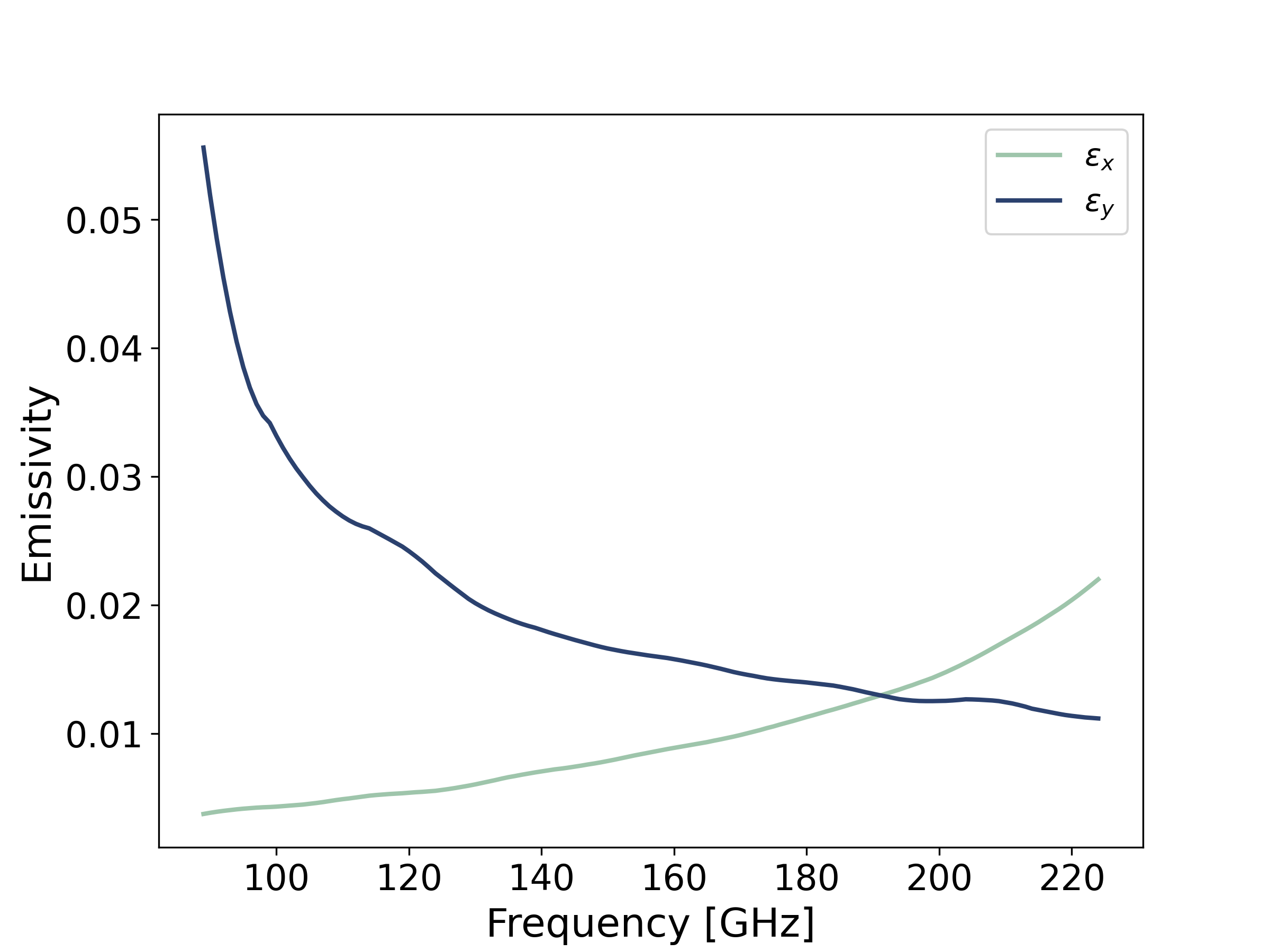}
\caption[]{HWP emission coefficients as a function of frequency for MFT frequency range, as described in \cite{LiteBIRDCollaboration:2023}. The blue and green lines refer to the capacitive and inductive axes, respectively.}
\label{fig:mft-hwp-emiss} 
\end{figure}

\sm{Figure \ref{fig:mft-hwp-emiss} shows the different values of emissivity along the two optical axes for a realistic HWP, in the frequency range covered by MFT. Note that this is shown here as an example of this behavior. In the analysis presented below, $A_{2f}$ is treated as a free parameter and explored over ranges motivated by the LiteBIRD HWP design and by the signal amplitudes derived in~\citep{Takaku:2023, Giardiello:2022, Micheli:2024}, assuming a 20-K HWP. This allows us to explore the systematic bias over the full frequency range while covering representative values of the HWPSS amplitude.}

We let the differential emissivity vary accordingly to compare with the requirements for non-linearity knowledge found in our previous work. In the presence of large synchronous signals, the detector response can be pushed out of the linear regime. The non-linear response of the TES then up-converts and couples the $2f$-HWPSS with the sky signal, leading to additional leakage terms in the map-making equations. The total \sm{leakage contribution to the TOD, arising from the combination of the HWP differential emission and the TES non-linear response,} becomes:
\begin{align}\label{eq:deltad2fnl}
\delta d_{NL}^{2f} &= \delta d_{NL} + \delta d^{2f} \notag \\
&\quad + g_1 \Big(
A_{2f}^2 \cos^2{\Theta} 
+ 2I A_{2f} \cos{\Theta} 
+ 2Q A_{2f} \cos{\Theta}\cos(2\chi) 
+ 2U A_{2f} \cos{\Theta}\sin(2\chi) \Big)
\end{align}
The full leakage maps, $\delta\hat{Q}$ and $\delta\hat{U}$, are then calculated by \sm{inserting} the mean non-linear signals $\langle \delta d_{\text{NL}}^{2f} \rangle$, $\langle \delta d_{\text{NL}}^{2f} \cos(2\chi) \rangle$ and $\langle \delta d_{\text{NL}}^{2f} \sin(2\chi) \rangle$ into eq.~\eqref{eq:mapmaking}. By analyzing the distribution of the cross-linking matrix terms for a given scanning strategy, we can often simplify the leakage equations by neglecting terms that are consistently very small, \sm{as shown in figure~\ref{fig:allhists}.} This allows for rapid generation of contaminated maps.

\subsection{Validation}
\begin{figure}[]
\begin{center}
    \includegraphics[width=0.9\textwidth]{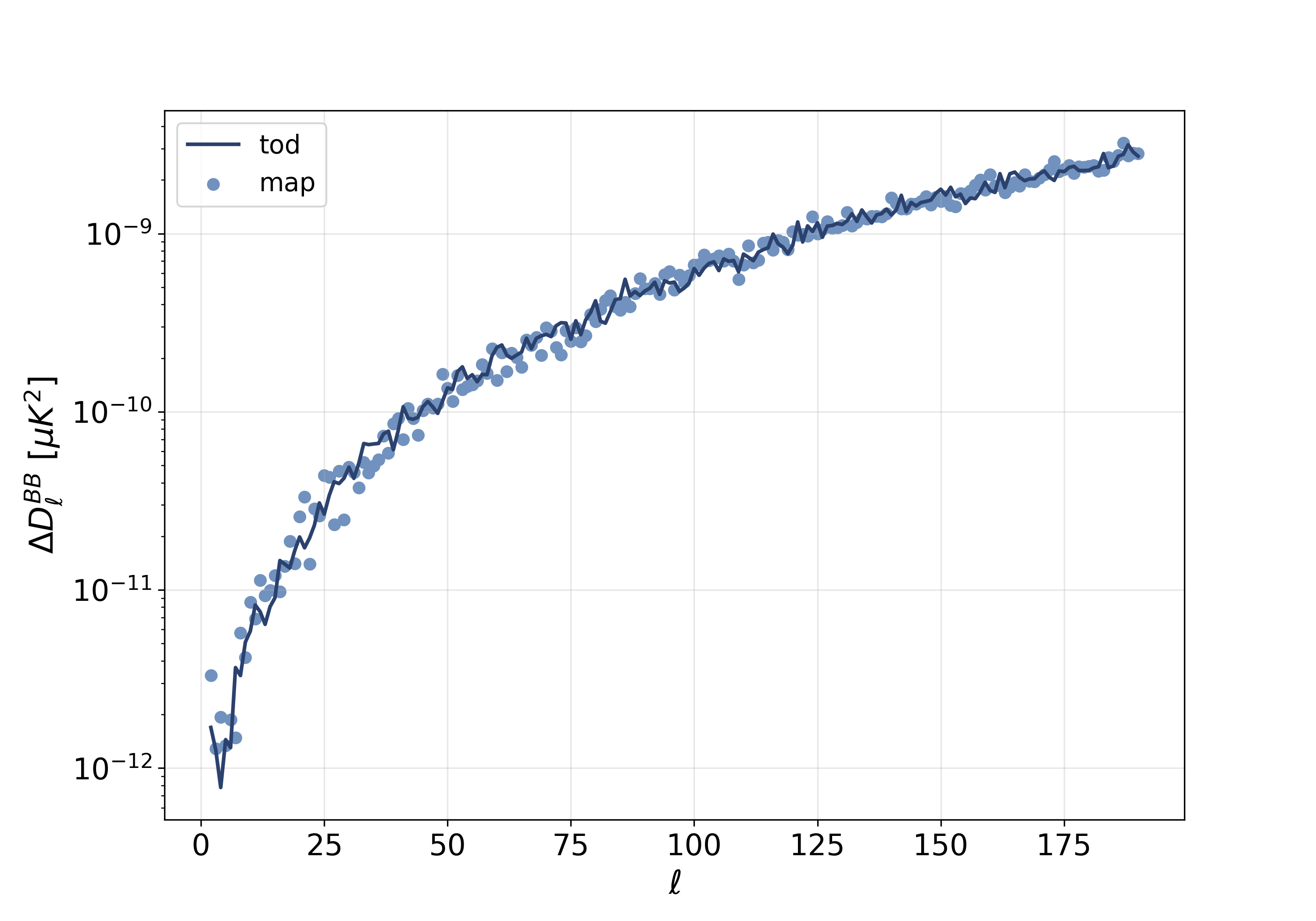}
\end{center}
\caption{\sm{\textit{BB} angular full-sky power spectra, $\Delta D_{\ell}$, of the difference between a simulated map with the systematic effect and a reference map without any systematic effects, isolating the leaked signal.} The solid line corresponds to the full TOD-based simulation, while the dots indicate the map-based approach. The panel demonstrates the consistency between the two methods.}\label{fig:spectra-comparison} 
\end{figure}

\begin{figure}[htbp]
\includegraphics[width=\textwidth]{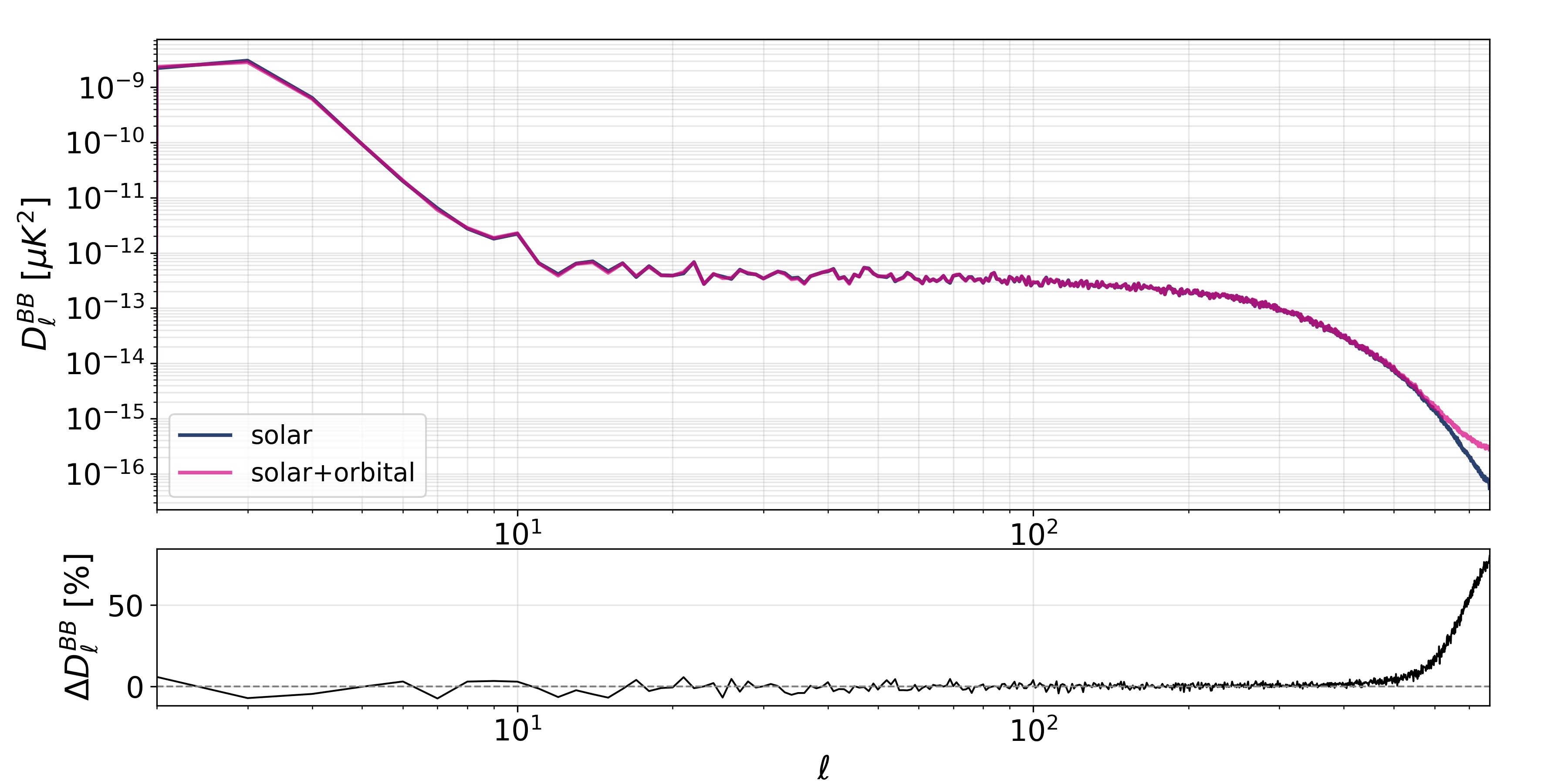}
\caption{Top panel: The $D_\ell^{BB}$ leakage power spectra for the solar-only (blue) and solar+orbital (pink) cases, in presence of nonlinearity. Bottom panel: Residuals between the two spectra.}\label{fig:Dipoles} 
\end{figure}

To validate this map-based formalism, we compare it against a full TOD simulation for a one-year, full-sky observation. The results show excellent agreement in the power spectra of the reconstructed leakage maps for both non-linearity alone and in combination with the HWPSS. This confirms that the map-based approach is a reliable and computationally efficient alternative to full TOD simulations. \sm{The formalism presented here does not account for non-orthogonal incidence of the incoming radiation, which is left to future extensions of this framework.} 
\newline
One limitation of this method is its inability to account for non-stationary signals, such as the orbital dipole signal, which changes as the spacecraft moves and cannot be captured by static cross-linking templates\footnote{Time-varying systematics could in principle be modeled by dividing the survey TOD into multiple time chunks, within which the systematic is assumed to remain stable but allowed to vary from one chunk to another \cite{McCallum:2022}.}. While the solar dipole can be included, neglecting the orbital dipole leads to a loss of power in the simulated leakage maps, as shown by comparing the two methods. \sm{This approximation has a minor impact on the reconstruction of the spectrum, as shown in figure \ref{fig:Dipoles}. The contribution of the orbital dipole may need to be handled separately in future studies.}
\newline
This formalism is particularly useful because it allows for speeding up the production of the contaminated maps, avoiding full TOD simulations. \sm{These involve iterating over the telescope pointing information, incurring a large computational cost, as this operation has to be repeated every time we change an instrumental parameter}. \sm{On the other hand, after the scanning strategy is set, with a map-based method we can calculate the cross-linking templates once and rapidly recover the output maps, and spectra, for several amplitudes of the systematic effect.} \newline \newline To perform the validation step, we consider a full-sky, one-year observation of a CMB-only realization. CMB maps are produced at nside=64 and smoothed
by Gaussian beam of the corresponding frequency channel (see Table 13 of \cite{LiteBIRDCollaboration:2023} for the corresponding FWHM). In particular, we consider two orthogonal detectors for the 140 GHz channel of \textit{LiteBIRD}. The TODs are produced using \texttt{litebird\_sim}\footnote{\url{https://github.com/litebird/litebird_sim/blob/master/litebird_sim/non_linearity.py}} \cite{Tomasi:2025}. The cross-link maps used in eq.~\eqref{eq:contStokes} are produced using the \texttt{Falcons} code\footnote{\url{https://github.com/yusuke-takase/Falcons.jl}}~\cite{Takase:2024}.
In figure~\ref{fig:spectra-comparison}, we show the \sm{residual power spectra between the input and the reconstructed output maps, in the case of the full TOD and map-based simulation}. The maps have been contaminated with an arbitrary value of non-linearity of 0.1~$\mathrm{K^{-1}}$ to validate the method. We see that the map-based approach captures well the features of the spectra, resulting in a reliable way to produce simulations of observations, including non-linearity from the detectors. In appendix~\ref{sec:appendixmaps}, we present the corresponding maps used in the computation of these angular power spectra, enabling a direct comparison between the two reconstruction methods.

\paragraph{Orbital dipole} As shown in previous studies,~\cite{Micheli:2024, Patanchon:2024}, a significant contribution to the total leakage comes from the dipole. Thus, it is important to account for its presence in the simulations. However, in the map-based approach, it is \sm{non-trivial} to include the contribution of the orbital dipole, being a non-stationary signal that is not projectable on maps. We provide an estimation of \sm{the effect of neglecting} the presence of the orbital dipole in figure~\ref{fig:Dipoles}, where we show the $BB$ spectra in two cases: with and without the orbital dipole for a TOD-based simulation. In this case, we have produced our maps at nside=256 to explore a wider range of multipoles. \sm{The residuals in the lower panel account for the discrepancy between the two cases. We can see how the difference becomes noticeable only at very high $\ell$s. Hence, this is not an issue at low multipoles, where the primordial $B$-modes we are interested in live.}

\section{Simulations and analysis}
\label{sec:simulations}

\subsection{Single frequency analysis}\label{sec:single-freq}
Having validated the formalism against full TOD simulations, we consider simplified scenarios where individual systematics can be isolated and studied. We begin with a single-frequency case to study the qualitative impact of TES non-linearity and HWP differential emission, first separately and then combined. We use this controlled setup to understand how each effect projects onto CMB residual maps, before moving to a multi-frequency analysis where the realistic impact on $r$ is assessed in the presence of foregrounds and after component separation.
We adopt the following procedure to determine the impact of the systematics on the reconstruction of the BB spectrum.
\begin{itemize}
    \item We build 100 realizations of CMB maps and white noise at $\texttt{nside}=64$, for the MFT M1-140~GHz band, \sm{considering a pair of orthogonal detectors. The noise is rescaled for the number of detectors.} These are generated from the \textit{Planck} fiducial cosmology using \texttt{CAMB}~\cite{Lewis:2011}, with parameters $A_s = 2\times10^{-9}$, $n_s = 0.965$, $\tau = 0.06$, and $r=0$, where $A_s$ and $n_s$ denote the amplitude and spectral index of the power spectrum of primordial scalar fluctuations, and $\tau$ is the reionization optical depth \cite{PlanckCollaboration:2020V}. The observed sky is obtained by coadding these maps with the solar dipole template and smoothing the resulting map with the angular resolution of a MFT M1-140~GHz detector. We chose to include the solar dipole signal, \sm{although its contribution is usually filtered with minimal impact on the final maps for an ideal instrument, because it contributes to the total signal that could drive a non-linear response in the detectors. Moreover, we are interested in studying how non-linearity can affect dipole reconstruction, as this could have an impact on the calibration strategy.} 
    
    \item We build 100 contaminated maps, considering different realizations of the amplitude of the systematic effect. More specifically, we extract $g_1$ from a normal distribution, with zero mean and $\sigma_{g_1}^2$ variance. We consider five different values of $\sigma_{g_1}^2$ and $A_{2f}$ to assess the impact on the reconstructed maps.
    
    \item For full sky simulations, we can use an exact likelihood calculation to derive the maximum likelihood value for the tensor-to-scalar ratio, for each realization \cite{Hamimeche:2008}:
    \begin{equation}
        \label{eq:likelihood}
        -2\ln \mathcal{L} (C_{\ell,obs}^{BB}|C_{\ell,mod}^{BB}) = \sum_{\ell=\ell_{min}}^{\ell=\ell_{max}} \left[ \frac{C_{\ell,obs}^{BB}}{C_{\ell,mod}^{BB}} - \ln{\frac{C_{\ell,obs}^{BB}}{C_{\ell,mod}^{BB}}} - 1 \right]
    \end{equation}
    with 
    \begin{equation}
        C_{\ell,mod}^{BB} = \frac{r}{r_0} C_{\ell}^{prim,r=r_0} + C_{\ell}^{lens} + N_{\ell}^{mod} \, ,
    \end{equation}
    \begin{equation}\label{eq:cl-obs}
        C_{\ell,obs}^{BB} = C_{\ell}^{lens} + C_{\ell}^{res}
    \end{equation}
    
    For the $i$-th realization, the noise model, $N_{\ell}^{mod}$, is calculated as the average over the $j \neq i$ residual noise realizations, to avoid correlations. We set $r_0 = 10^{-3}$ to avoid tilting effects of the spectrum. \sm{$C_{\ell}^{res}$ corresponds to the spectrum of the residuals map, obtained by subtracting the input CMB map spectrum from the total observed map.}
    
    \item We extract the maximum-likelihood value of the tensor to scalar ratio, for each realization, both with and without systematics. \sm{We refer to these quantities as $r_{\mathrm{sys}}$ and $r_{\mathrm{nosys}}$, respectively.} We define the bias on \textit{r} as $\delta r = r_{\mathrm{sys}}-r_{\mathrm{nosys}}$. 

    \item We calculate the average of the bias over the 100 simulations for each value of the systematics amplitude to recover the scaling and set the requirement, given the threshold of $6.5 \cdot 10^{-6}$, which corresponds to the error budget allocated for each systematic effect of 1\% of the targeted statistical uncertainty~\cite{LiteBIRDCollaboration:2023}. When considering a perturbation of a single channel, we rescale this budget by the number of total channels, namely $6.5 \cdot 10^{-6}/22$. In principle, different frequency channels may not contribute equally to the final requirement, meaning some channels could be more critical than others. In this work, we adopt an equal division of the total systematic budget among channels as a conservative assumption. As shown later in the results, high-frequency channels generally drive the requirement, implying that less stringent thresholds could be tolerated for certain low-frequency channels. Nevertheless, this uniform allocation allows us to explore a worst-case scenario without relying on assumptions about the relative importance of each channel, reducing the risk of underestimating residual effects that might become relevant even in channels expected to have a minor impact.
    
    We also consider the $\delta r$ distribution, as its standard deviation increases with the amplitude of the perturbation. This effect is observed in~\cite{Carralot:2025} for gain uncertainties, and we could use the same formalism to derive the requirement on non-linearity, eventually being a perturbation of the gain. Given that, we can define $\Delta(\delta r) = \sqrt{\bar{\delta r}^2 + \rm Var (\delta r)}$ to account simultaneously for both the mean value and the extra variance of the $\delta r$ distribution.
\end{itemize}
According to the formalism developed in section~\ref{sec:formalism}, we build and analyze the contaminated maps in three cases: 1) ideal HWP and non-ideal detector response, 2) non-ideal HWP and ideal detector response, and 3) non-ideal HWP and non-ideal detector response.

\begin{figure}
\includegraphics[width=\textwidth]{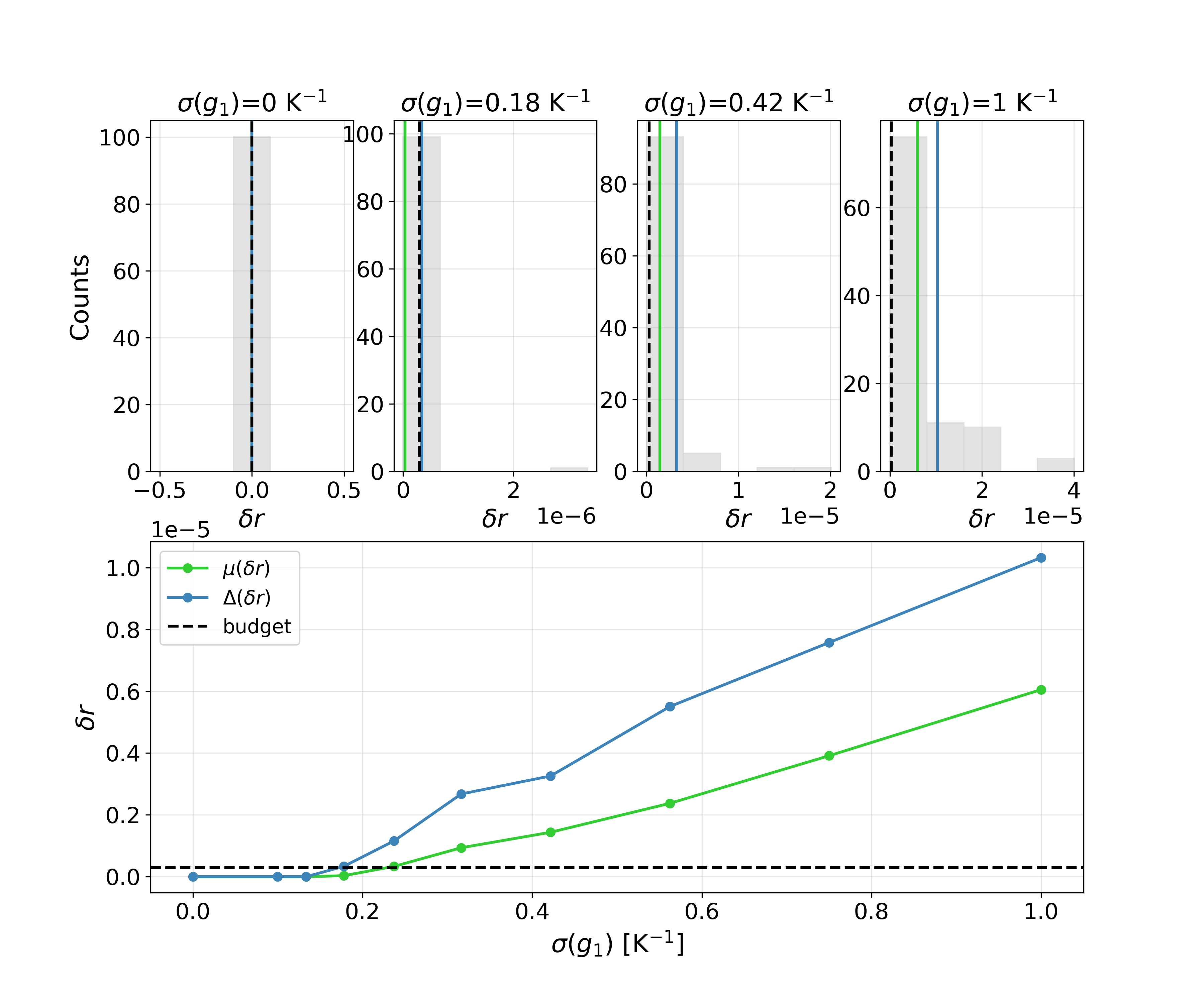}
\caption[Distribution of the bias on $r$ over 100 realizations versus $\sigma(g_1)$ in the absence of HWPSS, together with its mean and RMS.]{\textit{Top}: Distribution over 100 realizations of the bias on $r$ for different levels of non-linearity knowledge $\sigma(g_1)$ in the absence of HWPSS. Vertical lines indicate the mean (green) and RMS (blue) of each distribution, while the dashed black line represents the channel systematics budget, $6.5 \cdot 10^{-6}/22$. \textit{Bottom}: Mean and RMS of the $\delta r$ distributions as a function of $\sigma(g_1)$. The same color code is applied.}
\label{fig:bias_from_g} 
\end{figure}

\paragraph{Ideal HWP and non-ideal detector response}  
When only TES non-linearity is present and the HWP is ideal, some leakage arises solely from quadratic couplings of the CMB signal and the dipole.\footnote{White noise is coupled as well; however, its impact is minimal. If foregrounds are present, they also couple similarly and can be accounted for by simply coadding them to the input maps. The treatment is analogous and used in the following.} In this regime, the amplitude of the residuals' spectra, defined as the difference between the contaminated maps (with CMB and white noise as input) and the input CMB, grows monotonically with $\sigma_{g1}$. This case is mainly conceptual: in the absence of large signals such as the HWP emission, the TES essentially operates in its linear regime. The corresponding bias on $r$, calculated following the procedure described above and shown in figure~\ref{fig:bias_from_g}, remains negligible except for unrealistically large values of $\sigma_{g1}$, given that we expect a typical \lb\ detector to have $g_1 \simeq -0.144~\mathrm{K^{-1}}$ from previous work\footnote{In \cite{deHaan:2024spie}, the typical non-linearity is expressed in inverse power units, with an absolute value of $0.8\,\mathrm{pW^{-1}}$. This value is converted following the procedure described in section~3.1 of \cite{Micheli:2024}.
}.
Both the mean and RMS of the $\delta r$ distributions are shown for comparison with the channel budget, represented by the black dashed line.

\begin{figure}
\includegraphics[width=\textwidth]{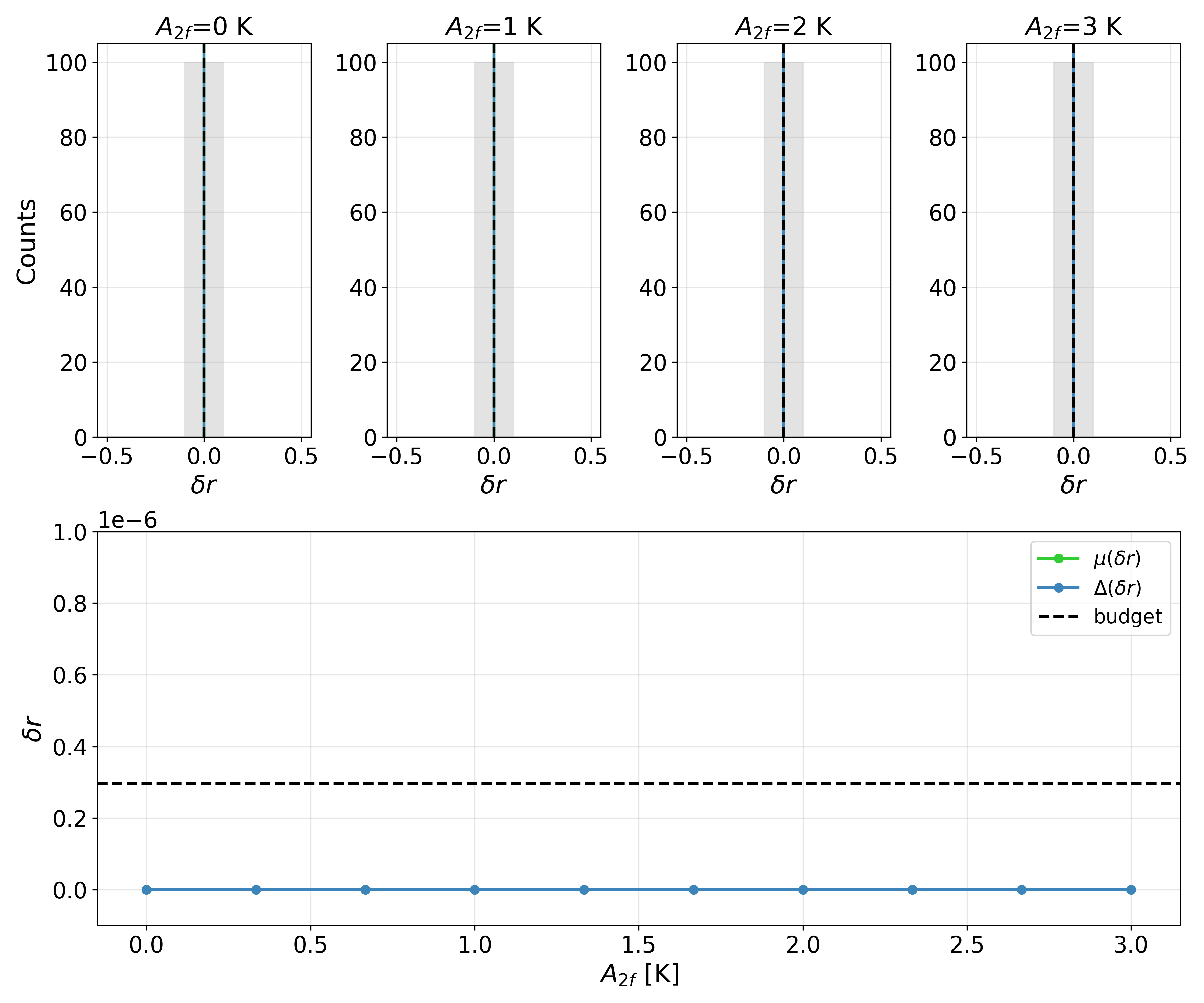}
\caption[Distribution of the bias on $r$ over 100 realizations versus $A_{2f}$ at $\sigma(g_1)=0$, together with its mean and RMS.]{\textit{Top}: Distribution over 100 realizations of the bias on $r$ for different levels of HWP 2f amplitude $A_{2f}$ at fixed $\sigma(g_1)=0$. Vertical lines indicate the mean (green) and RMS (blue) of each distribution, while the dashed black line represents the channel systematics budget. \textit{Bottom}: Mean and RMS of the $\delta r$ distributions as a function of $A_{2f}$. The bias on $r$ is consistently null. The same color code as in the upper panel is applied.}
\label{fig:bias_from_a} 
\end{figure}

\paragraph{Non-ideal HWP and ideal detector response}  
Conversely, if the TES response is perfectly linear but the HWP is non-ideal, symmetry ensures that leakage terms vanish in the case of an ideal scanning strategy. In practice, imperfect sky sampling leaves residuals that project the HWPSS appearing at $2f$ onto the sky, acting as an additional noise-like term which scales with the injected amplitude $A_{2f}$. As expected, since the differential emissivity signal is purely additive and does not modulate the sky (as would do, for example, differential transmission), the corresponding bias on $r$ is negligible, as shown in figure~\ref{fig:bias_from_a}. 

\begin{figure}
\centering
\includegraphics[width=\textwidth]{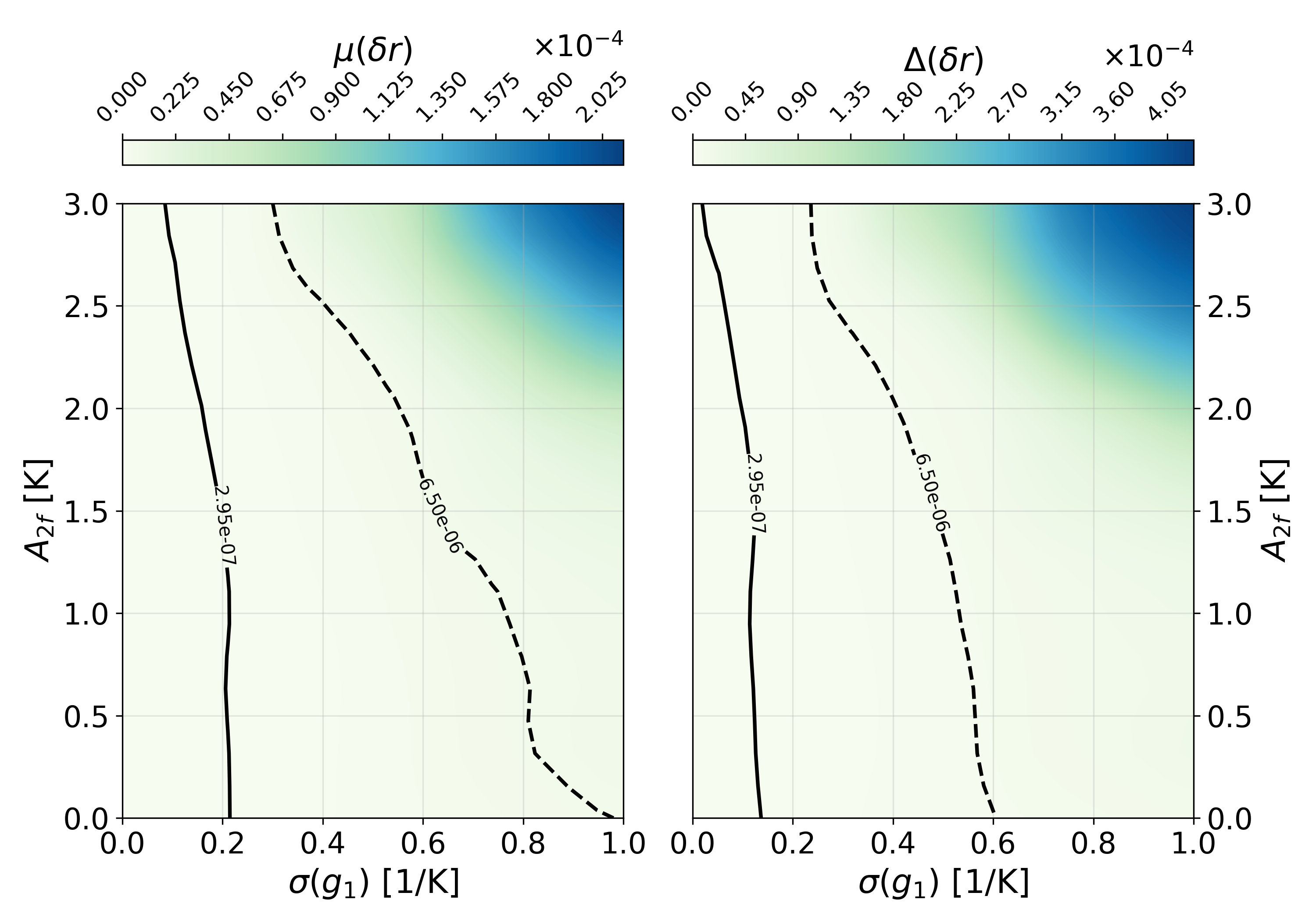}
\caption[Contour plot of the mean and RMS of the bias on $r$ over 100 realizations versus $A_{2f}$ and $\sigma(g_1)$.]{\textit{Left}: Contour plot of the mean $\delta r$ over 100 Monte Carlo realizations as a function of $g_1$ and $A_{2f}$. The input maps contain the CMB, the solar dipole, and a white noise realization at 140~GHz. The black solid line indicates the rescaled requirement for a single frequency channel, while the black dashed line marks the overall systematic error budget. \textit{Right}: Same as left, but showing the RMS of $\delta r$.}
\label{fig:contour-cmb} 
\end{figure}

\paragraph{Non-ideal HWP and non-ideal detector response}  
When both $\sigma(g_1)$ and $A_{2f}$ are nonzero, their interaction produces additional leakage, most notably through $2f \rightarrow 4f$ terms. \sm{From eq.~\eqref{eq:deltad2fnl}, we can notice how a generic sinusoidal input at frequency $\omega$ generates a component at $2\omega$ when TES non-linearity is present. As a result, in the presence of a HWPSS at $2f_{\mathrm{HWP}}$, detector non-linearity leads to the appearance of spurious contributions that leak into the science band.}
Figure~\ref{fig:contour-cmb} shows the mean and RMS of the $\delta r$ distribution over 100 Monte Carlo realizations as a function of $\sigma(g_1)$ and $A_{2f}$. As expected, the bias grows with increasing $\sigma(g_1)$ and $A_{2f}$. These contour plots identify the region of parameter space where the combined systematic remains below the channel budget, thus providing a joint requirement on both parameters. The black solid line marks the channel budget threshold, with points to the left being acceptable. % For comparison, we also consider a dipole-free configuration. While non-physical, this case is instructive, as it clearly illustrates the significant role of dipole coupling. As shown in Figure~\ref{fig:contour-cmb-nodip}, removing the dipole drastically reduces its impact on $r$, highlighting the importance of accounting for it before foreground cleaning and further analysis.
This exercise also demonstrates the advantage of our map-based framework. \sm{The generation of $100\times10\times10$ maps, corresponding to 10 values of $\sigma(g_1)$, 10 values of $A_{2f}$, and 100 CMB and noise realizations, for one year of observations with two detectors requires only about three minutes on a standard laptop. This is achieved after a one-time computation of the cross-linking matrix, which requires about 1 hour.} In contrast, generating the same number of full TOD simulations would demand approximately 24 CPU-hours\footnote{The time-domain simulationa have been run on the CINECA Galileo100 cluster. Further technical specifications are available in the CINECA HPC Documentation: \url{https://docs.hpc.cineca.it/hpc/galileo.html}.}. This efficiency makes it possible to explore a vast parameter space in a reasonable amount of time. The map-based pipeline is also far more efficient in terms of memory: it can be run comfortably on a laptop, whereas producing long TODs for many detectors quickly becomes unfeasible. 

\subsection{Multi-frequency analysis}

\begin{figure}[]
    \centering
    \includegraphics[width=\linewidth]{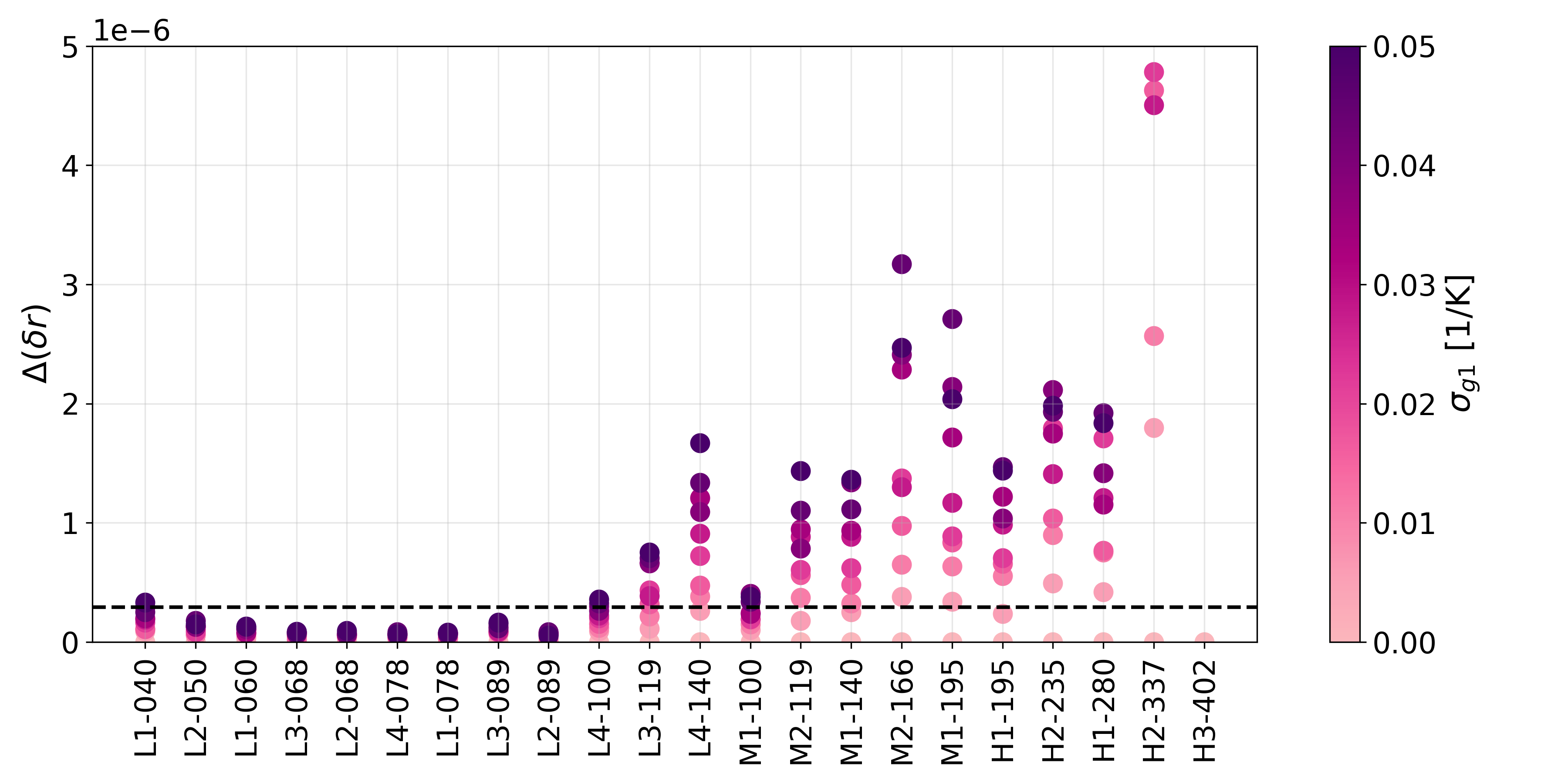}
    \caption[RMS of the $\delta r$ bias over 100 realizations as a function of non-linearity knowledge $\sigma(g_1)$, for all \lb\ frequency channels.]{RMS of the $\delta r$ bias over 100 realizations as a function of non-linearity knowledge $\sigma(g_1)$ in absence of HWPSS, for all \lb\ frequency channels. Input maps include CMB, the solar dipole, white noise, and the \texttt{d0s0} foreground model for dust and synchrotron emission, as implemented in the \texttt{PySM} package \citep{Panexp:2025}. The black dashed line marks the per-channel allocation of the channel systematic error budget, $6.5\cdot10^{-6}/22$.}
    \label{fig:bias-sigmag-allchannels}
\end{figure}

\begin{figure}[]
    \centering
    \includegraphics[width=\linewidth]{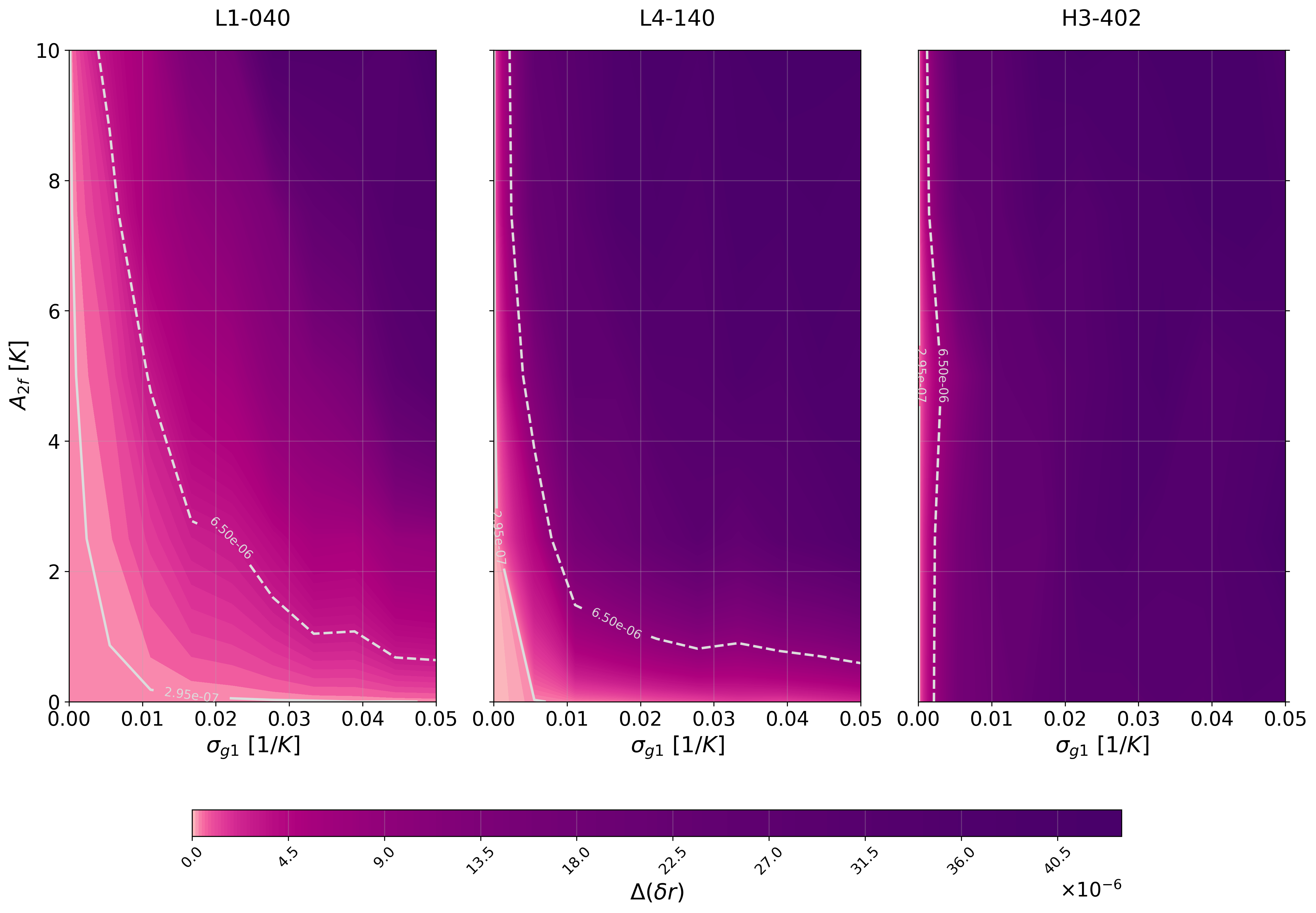}
    \caption[]{Contour plots of the RMS of the $\delta r$ distribution over 100 Monte Carlo realizations, for three representative \lb\ channels. Input maps include CMB, the solar dipole, white noise, and the \texttt{d0s0} foreground model for dust and synchrotron emission, as implemented in the \texttt{PySM} package \citep{Panexp:2025}. Each channel is perturbed independently by adding HWPSS and non-linearity. The white solid line marks the rescaled single-channel requirement, and the dashed line indicates the total systematic error budget.}
    \label{fig:bias-joint-d0s0}
\end{figure}

\begin{figure}[]
    \centering
    \includegraphics[width=\linewidth]{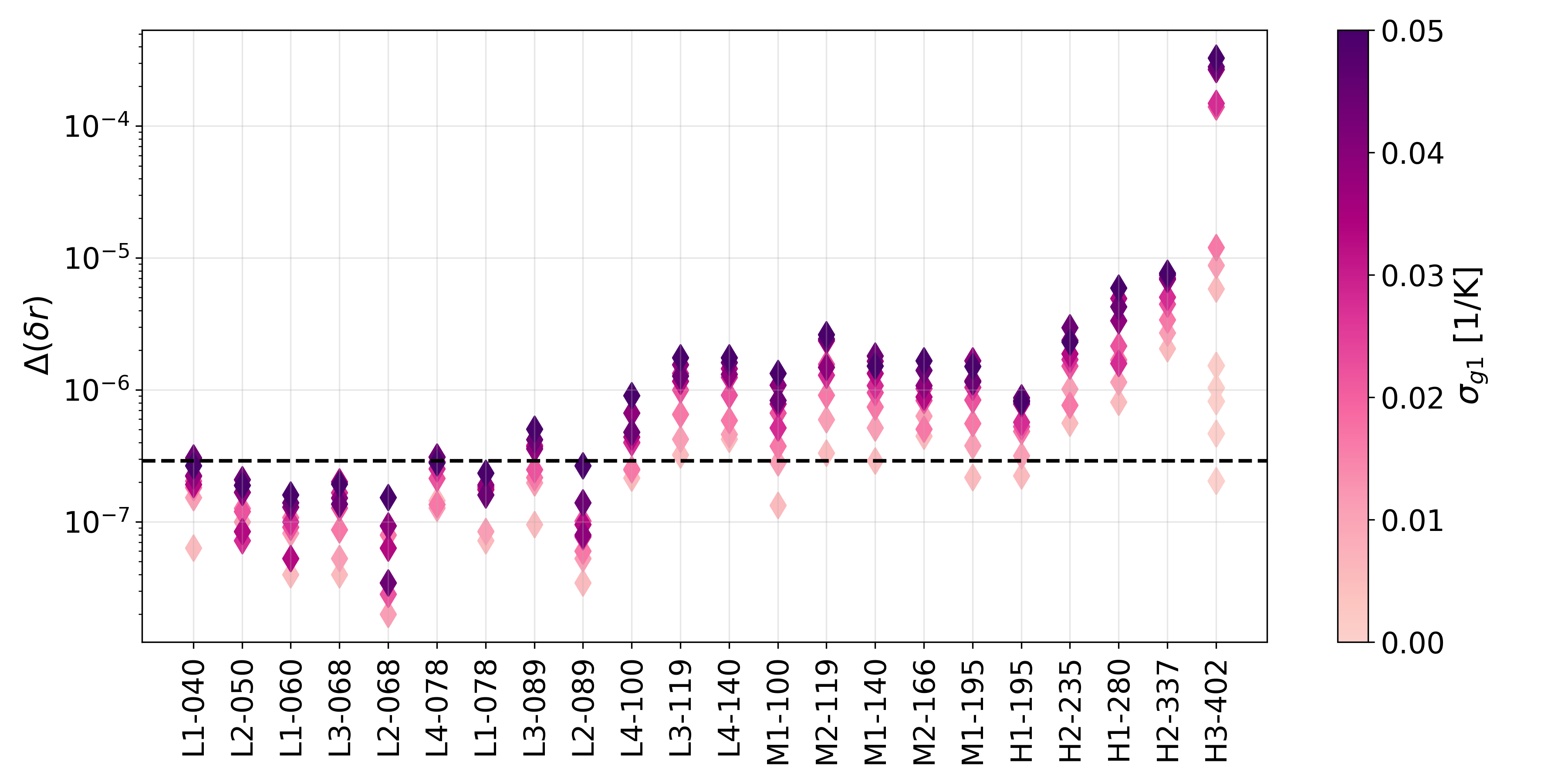}
    \caption[RMS of the $\delta r$ bias over 100 realizations as a function of non-linearity knowledge $\sigma(g_1)$, for all \lb\ frequency channels.]{RMS of the $\delta r$ bias over 100 realizations as a function of non-linearity knowledge $\sigma(g_1)$ in absence of HWPSS, for all \lb\ frequency channels. Input maps include CMB, the solar dipole, white noise, and the \texttt{d1s1} foreground model for dust and synchrotron emission, as implemented in the \texttt{PySM} package \citep{Panexp:2025}. The black dashed line marks the per-channel allocation of the overall systematic error budget, $6.5\cdot10^{-6}/22$.}
    \label{fig:bias-sigmag-allchannels_d1s1}
\end{figure}

\begin{figure}[]
    \centering
    \includegraphics[width=\linewidth]{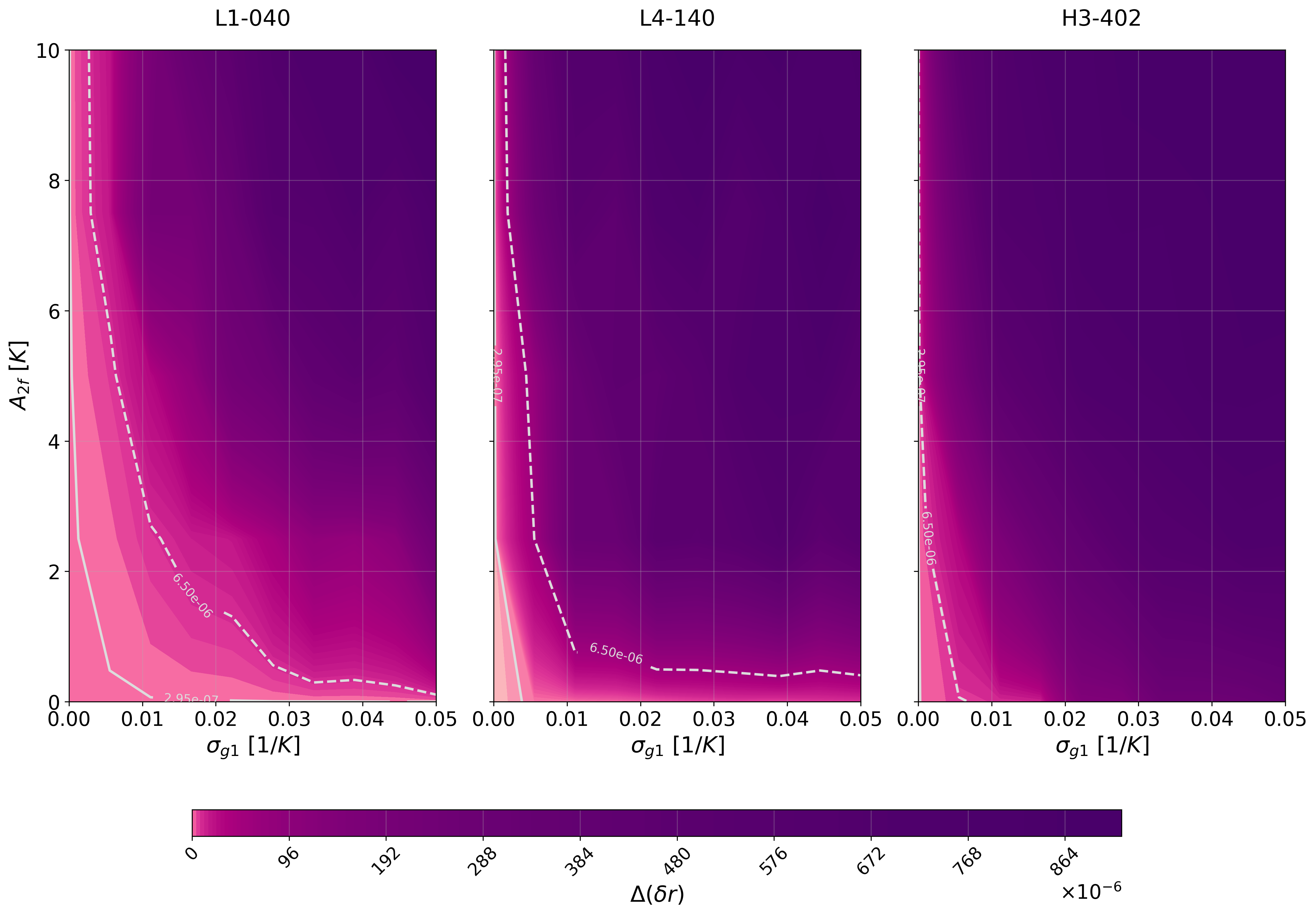}
    \caption[]{Contour plots of the RMS of the $\delta r$ distribution over 100 Monte Carlo realizations, for three representative \lb\ channels. Input maps include CMB, the solar dipole, white noise, and the \texttt{d1s1} foreground model. Each channel is perturbed independently by adding HWPSS and non-linearity. The white solid line marks the rescaled single-channel requirement, and the dashed line indicates the total systematic error budget.}
    \label{fig:bias-joint-d1s1}
\end{figure}

We extend our analysis to multi-frequency simulations to assess the impact of foregrounds (and foreground cleaning) in the presence of coupled instrumental systematics. As before, the goal is to quantify the residual bias on the recovered CMB signal and identify \sm{which frequency channels are most critically impacted by the systematic effect}.
The analysis pipeline proceeds as follows. For each of the 100 Monte Carlo realizations:  
\begin{enumerate}
    \item We generate input sky maps including the CMB, white noise, the solar dipole, and the polarized Galactic foregrounds. \sm{We adopt the 
    \texttt{d0s0} foreground model for dust and synchrotron emission, as implemented in the \texttt{PySM} package \citep{Panexp:2025}.}
    
    \item Each frequency channel is independently perturbed by injecting both a HWPSS at 2$f$ and detector non-linearity. \sm{We explore a grid of parameters spanning 0 to 0.02~$\mathrm{K^{-1}}$ (in 10 uniform steps) for $\sigma_{g1}$, and 0 to 10~$\mathrm{K}$ (in 5 uniform steps) for $A_{2f}$, as summarized in figures \ref{fig:bias-joint-d0s0} and \ref{fig:bias-joint-d1s1}.}
    
    \item \sm{The perturbed maps are processed through a blind foreground cleaning pipeline, employing a blind component separation method to recover the CMB signal. In particular, we adopt the Needlet Internal Linear Combination (NILC) algorithm \cite{Bennett:2003a, Delabrouille:2009}.} Before component separation, all frequency maps are smoothed to a common resolution of 70.5~arcmin, corresponding to the lowest angular resolution of \lb. 
    
    \item To compute angular power spectra, we apply a common Galactic mask corresponding to the \texttt{GAL60} \textit{Planck} mask\footnote{\url{https://irsa.ipac.caltech.edu/data/Planck/release_2/ancillary-data/previews/HFI_Mask_GalPlane-apo5_2048_R2.00/index.html}}, with an additional 10\% exclusion. The extra cut is derived by smoothing the mean residual map with a $3^\circ$ Gaussian beam, ranking pixels by absolute amplitude, and masking the 10\% most contaminated ones. This yields an effective sky fraction of $f_{\mathrm{sky}}=0.5$, consistent with the masking strategy adopted in \cite{LiteBIRDCollaboration:2023}.
    
    \item We estimate the maximum-likelihood value of $r$ for both the ideal and perturbed cases from eq.~\eqref{eq:likelihood}. Here, we replace the model noise spectrum with a model foreground and noise residual spectrum, calculated as the average over the $ j \neq i$ NILC output residual spectra of noise and foregrounds. The spectra $C_{\ell}^{res}$ in eq.~\eqref{eq:cl-obs} are computed for the residual CMB maps, obtained by subtracting the input CMB $B$-mode map from the total observed $B$-mode map recovered by NILC.
    
    \item Finally, we derive the distribution of the bias and extract the mean value and the RMS for each realization of the systematic effects.
\end{enumerate}
This procedure is repeated independently for each frequency channel, enabling us to evaluate if and how component separation deals with different types of contamination, and to determine which bands are most critical in setting the systematic requirements. This is particularly relevant in the case of coupled systematics, where the degeneracy between them can mask their individual contributions. Even without HWPSS, the non-linear detector response can induce a spurious polarization signal, especially in high-frequency bands where foreground emission is strong and the direct I-Q mixing is non-negligible. This effect is particularly evident in the HFT H3-402~GHz channel, which yields the tightest constraint on $\sigma_{g1}$, namely $\sigma_{g1} < 1.3 \times 10^{-4}\,\mathrm{K}^{-1}$ for the single channel. This threshold is derived from the best-fit relation between $\delta r$ and $\sigma_{g1}$ shown in figure~\ref{fig:bias-sigmag-allchannels}. \sm{To improve the reliability of the fit, given the sparse sampling of this channel, we performed a dedicated set of additional simulations exploring the range between 0.0001 and 0.005 $\mathrm{K^{-1}}$ for $\sigma_{g1}$.}
After exploring the full parameter space, we visualize the allowed regions in the $(A_{2f},\,\sigma_{g1})$ plane for representative frequency bands, as shown in figure~\ref{fig:bias-joint-d0s0}. Consistent with the single-frequency analysis, the highest-frequency channels dominate the requirements. In these bands, residual foreground leakage amplifies the bias more than the direct coupling between non-linearity and HWPSS, indicating that component separation efficiency plays a central role in shaping the final performance. Conversely, the lower-frequency channels are less affected, leading to more relaxed requirements on non-linearity and HWPSS amplitude. \sm{From these analyses, we find that at low and intermediate frequencies the control of the HWPSS amplitude is a key element in the trade-off with detector non-linearity (see the left panel of figure~\ref{fig:bias-joint-d0s0}). In this regime, the HWP design should therefore aim at minimizing the imbalance between the two optical axes, which directly sets the level of intensity-to-polarization leakage. At higher frequencies, a different criterion can be applied. As shown in figure~\ref{fig:bias-sigmag-allchannels}, detector non-linearity alone introduces a non-negligible bias even in the absence of HWPSS, due to the leakage from Galactic foregrounds. In this case, requiring a vanishing optical-axis imbalance is neither necessary nor meaningful. Instead, a relevant requirement could be to balance the HWP emissivity such that the resulting differential emission remains subdominant with respect to Galactic foregrounds. Future work will focus on the development of HWP models explicitly incorporating these frequency-dependent design criteria.}
In this part of the analysis, we adopted a simplified sky model, since the combined presence of multiple instrumental effects already introduces a complex set of couplings. This controlled setup allows us to isolate the main dependencies of the systematic bias and to establish a quantitative baseline for more realistic simulations. To test the robustness of these conclusions, we also analyzed a more realistic sky configuration using the \texttt{PySM} \texttt{d1s1} model, which includes spatial variability of the dust and synchrotron spectral properties \cite{Panexp:2025}. The component separation is performed applying a Multiclustering needlet ILC (MC-NILC), which is particularly suitable for complex sky models \cite{Carones:2023}.
We repeated the same analysis as for the \texttt{d0s0} sky. The results are qualitatively similar: in the case of non-linearity only, the requirement is still driven by the highest-frequency channel. We obtain a requirement of $\sigma_{g1} < 1.2 \times 10^{-4} \, \mathrm{K}^{-1}$ for the non-linearity only case, which is reported in figure \ref{fig:bias-sigmag-allchannels_d1s1}. The joint allowed regions in the $(A_{2f},\,\sigma_{g1})$ plane for three representative frequency channels are shown in figure~\ref{fig:bias-joint-d1s1}. Table \ref{tab:summary-results} shows a summary of the results of this work, providing the threshold value of non-linearity knowledge for the three representative bands, both in the absence of HWPSS at 2$f$, and for some typical amplitudes of it \cite{Micheli:2024}. The feasibility of these requirements clearly depends on both the HWP design and the detector calibration strategy. For typical values of the TES non-linearity, of order $10^{-2},\mathrm{K}^{-1}$, the requirement at 140~GHz corresponds to a relative accuracy of about $20\%$. The most stringent requirements at the highest frequencies corresponds to a relative accuracy at the percent level. These values provide useful targets for dedicated calibration efforts. The TES response can eventually be reconstructed following the method presented in \cite{deHaan:2024}, while dedicated ground-based calibration measurements could further improve the knowledge of the TES non-linearity, particularly for the highest-frequency channels. In parallel, ongoing studies are investigating the optimization of the HWP optical properties to reduce differential emissivity and hence the HWPSS amplitude.

\begin{table}[h!]
    \centering
    \begin{tabular}{c c c c}
        \toprule
        Frequency [GHz] & Sky model &
        \makecell{$\sigma_{g1}^{\mathrm{req}}$ [K$^{-1}$] \\ ($A_{2f} = 0$ K)} &
        \makecell{$\sigma_{g1}^{\mathrm{req}}$ [K$^{-1}$] \\ ($A_{2f}^{\nu} \neq 0$)} \\
        \midrule
        
        40  & d0s0 & $4.5 \times 10^{-2}$ & $4.5 \times 10^{-2}$ ($A_{2f}^{\nu} = 0$ K) \\
        40  & d1s1 & $4.7 \times 10^{-2}$ & $4.7 \times 10^{-2}$  ($A_{2f}^{\nu} = 0$ K)  \\
        140 & d0s0 & $8.4 \times 10^{-3}$ & $3.3 \times 10^{-3}$  ($A_{2f}^{\nu} = 0.2$ K)  \\
        140 & d1s1 & $5.7 \times 10^{-3}$ & $ 1.8 \times 10^{-3}$  ($A_{2f}^{\nu} = 0.2$ K) \\
        402 & d0s0 &  $1.3 \times 10^{-4}$ & {$9.9 \times 10^{-5}$} ($A_{2f}^{\nu} = 1.3$ K) \\
        402 & d1s1 &  $1.2 \times 10^{-4}$ & {$9.9 \times 10^{-5}$} ($A_{2f}^{\nu} = 1.3$ K) \\
        
        \bottomrule
    \end{tabular}
    \caption{Summary of the results of this work. Three representative bands are reported, for two sky models with different complexities. We show the requirements on the level of knowledge of non-linearity for each band, both in the absence of HWPSS at 2$f$ and for typical values of it at the corresponding frequencies, $A_{2f}^{\nu}$, when different from zero.}
    \label{tab:summary-results}
\end{table}

\begin{comment}
\sm{ \section{New HWP models}
We can compare these results with some new HWP models under study. The plots below show the respective impact in terms of differential emissivity, and the corresponding values in the parameter space for the requirements. This is the case for a d1s1 sky for three representative channels.}
\begin{figure}[]
    \centering
    \includegraphics[width=\linewidth]{figs-from-thesis/Screenshot 2025-11-19 at 11.52.13.png}
    \caption{}
    \label{}
\end{figure}

\begin{figure}[]
    \centering
    \includegraphics[width=0.6\linewidth]{figs-from-thesis/deps-vs-hwpmodels.png}
    \caption{}
    \label{}
\end{figure}
\end{comment}

\section{Conclusions}
\label{sec:conclusions}

In this work, we have developed and validated a fast, map-based formalism to simulate and quantify the impact of TES detector nonlinearity and HWP non-idealities on CMB polarization measurements. This method directly models the leakage these systematics induce at the map level, bypassing the need for computationally expensive time-domain simulations. Comparison with full TOD simulations demonstrates that the formalism accurately captures the relevant physical effects while reducing computational costs by several orders of magnitude.
We have applied this framework to the specific case of the \textit{LiteBIRD} mission, adopting its baseline scanning strategy and frequency configuration as a realistic testbed. Our analysis isolates the impact of each systematic effect, before studying their coupling. Individually, both effects generate negligible bias on the recovered tensor-to-scalar ratio, provided their amplitudes remain within the expected instrumental range. However, when combined, their interaction leads to non-trivial leakage terms, producing a measurable bias (see appendix \ref{sec:appendix} for the full derivation). The coupling is particularly relevant in the presence of large signals such as the solar dipole, confirming previous findings that dipole-synchronous signals can drive spurious polarization leakage if not correctly accounted for.
The map-based framework allows for extensive Monte Carlo exploration of the parameter space, enabling the derivation of joint requirements for both non-linearity and HWP synchronous signal amplitude. Multi-frequency analyses, including a full treatment of Galactic foregrounds via blind component separation, confirm that these requirements are robust: the dominant constraints arise from high-frequency channels, where foregrounds are brighter and I-P leakage is stronger. Conversely, low-frequency bands are less sensitive to the coupling, offering a higher margin on detector non-linearity. Overall, our results demonstrate that the interplay between detector non-linearity and HWP differential emission constitutes a potentially relevant source of bias for next-generation CMB polarization missions targeting $r \lesssim 10^{-3}$. The formalism developed here provides a practical and efficient tool to evaluate such effects, complementing detailed end-to-end simulations. 

\appendix
\section{Full derivation of the map-based formalism}
\label{sec:appendix}

For ease of reference, we rewrite here the binner map-maker equation:
\begin{equation}\label{eq:app-binner}
{\hat{\textbf{S}}}=
\begin{pmatrix}
\hat{I} \\
\hat{Q} \\
\hat{U}
\end{pmatrix}
= M^{-1}
\begin{pmatrix}
\langle d_j \rangle \\
\langle d_j \cos (2\chi_j) \rangle \\
\langle d_j \sin (2\chi_j) \rangle
\end{pmatrix},
\end{equation}
where $M$ is the cross-linking matrix:
\begin{equation}
M = \begin{pmatrix}
1 & \langle\cos (2\chi_j) \rangle & \langle\sin (2\chi_j) \rangle \\
\langle\cos (2\chi_j) \rangle & \langle\cos^2 (2\chi_j) \rangle  & \langle\cos (2\chi_j)\sin(2\chi_j) \rangle\\
\langle\sin (2\chi_j) \rangle & \langle\sin (2\chi_j)\cos (2\chi_j) \rangle & \langle\sin^2 (2\chi_j) \rangle
\end{pmatrix}.
\end{equation}
To keep the formalism as general as possible, we redefine the entries of $M$ as:
\begin{equation}
    \begin{aligned}
        A &= \langle\cos{2\chi} \rangle, & 
        B &= \langle\sin{2\chi} \rangle, \\
        C &= \langle\cos^2{2\chi} \rangle, &
        D &= \langle\cos{2\chi}\sin{2\chi} \rangle, \\
        E &= \langle\sin^2{2\chi} \rangle, &
    \end{aligned}
\end{equation}
where, $2\chi=4\theta-2\psi$ and we have dropped the $j$ subscript to simplify the notation. With these choices, the cross-link matrix reads
\begin{equation}
M = 
    \begin{pmatrix}
        1 & A & B \\
        A & C & D \\
        B & D & E 
    \end{pmatrix}.
\end{equation}
and its inverse as
\begin{equation}
M^{-1} = \frac{1}{\det M}
    \begin{pmatrix}
        CE-D^2 & BD-AE & AD-BC \\
        BD-AE & E-B^2 & AB-D \\
        AD-BC & AB-D & C-A^2 
    \end{pmatrix},
\end{equation}
where
\begin{equation}
    \det M = CE - D^2 - A(AE-BD) + B(AD-BC).
\end{equation}
We now compute the right-hand side of eq.~\eqref{eq:app-binner} when the signal is perturbed by detector non-linearity. Denoting the contaminated TOD as $d_{\mathrm{NL}}$, 
\begin{equation}
\label{eq:pixav-nl-only}
    \begin{cases}
        \langle d_{NL} \rangle = I + A Q + B U 
        + g_1\big(I^2 + C Q^2 + E U^2 + 2A I Q + 2B I U + 2D Q U \big) \\[4pt]
        \langle d_{NL}\cos{2\chi} \rangle = A I + C Q + D U \\
        \qquad\qquad\qquad\;
        + g_1\big(A I^2 + F Q^2 + G U^2 + 2C I Q + 2D I U + 2H Q U\big) \\[4pt]
        \langle d_{NL}\sin{2\chi} \rangle = B I + D Q + E U \\
        \qquad\qquad\qquad\;
        + g_1\big(B I^2 + H Q^2 + J U^2 + 2D I Q + 2E I U + 2G Q U\big)
    \end{cases}
\end{equation}
Here, we introduced the higher-order spin-weighted quantities:
\begin{equation}
    \begin{aligned}
        F &= \langle\cos^3{2\chi} \rangle, & 
        G &= \langle\sin^2{2\chi}\cos{2\chi} \rangle, \\
        H &= \langle\sin{2\chi}\cos^2{2\chi} \rangle, & 
        J &= \langle\sin^3{2\chi} \rangle .
    \end{aligned}
\end{equation}
Plugging eqs.~\eqref{eq:pixav-nl-only} into eq.~\eqref{eq:app-binner} gives the estimated Stokes vector including the systematic effect:
\begin{equation}
\label{eq:contStokes}
\hat{\textbf{S}} = \frac{1}{\det M}
    \begin{pmatrix}
        (CE-D^2) \langle d_{NL} \rangle + (BD-AE) \langle d_{NL}\cos{2\chi} \rangle + (AD-BC) \langle d_{NL}\sin{2\chi} \rangle\\[4pt]    
        (BD-AE) \langle d_{NL} \rangle + (E-B^2) \langle d_{NL}\cos{2\chi} \rangle + (AB-D) \langle d_{NL}\sin{2\chi} \rangle\\[4pt]
        (AD-BC) \langle d_{NL} \rangle + (AB-D) \langle d_{NL}\cos{2\chi} \rangle + (C-A^2) \langle d_{NL}\sin{2\chi} \rangle
    \end{pmatrix} 
\end{equation}
By explicitly evaluating $\hat{\textbf{S}}$, we derive the following leakage maps for $Q$ and $U$:
\begin{equation}\label{eq:leakQ-nl}
    \begin{split}
        \delta \hat{Q} \equiv \hat{Q} - Q = & \; \frac{g_1}{\det M} \Big[(BD-AE)\big(I^2+CQ^2+2AIQ+2BIU+2DQU\big) \\ 
        & \qquad + (E-B^2)\big(AI^2+FQ^2+GU^2+2CIQ+2DIU+2HQU\big) \\
        & \qquad + (AB-D)\big(BI^2+HQ^2+DU^2+2DIQ+2EIU+2GUQ\big)\Big],
    \end{split}
\end{equation}
\begin{equation}\label{eq:leakU-nl}
    \begin{split}
       \delta \hat{U} \equiv \hat{U} - U = & \; \frac{g_1}{\det M} \Big[(AD-BC)\big(I^2+CQ^2+2AIQ+2BIU+2DQU\big) \\ 
       & \qquad + (AB-D)\big(AI^2+FQ^2+GU^2+2CIQ+2DIU+2HQU\big) \\ 
       & \qquad + (C-A^2)\big(BI^2+HQ^2+DU^2+2DIQ+2EIU+2GUQ\big)\Big].
    \end{split}
\end{equation}
As expected, the leakage vanishes when $g_1\to 0$. eqs.~\eqref{eq:leakQ-nl} and \eqref{eq:leakU-nl} show that, once the spin-weighted terms are known, leakage maps can be estimated rapidly for a scan of the non-linearity parameter space by rescaling for $g_1$. An \textit{ideal} scanning strategy would let only a few terms to survive \cite{Couchot:1999}, given $\langle\cos{2\chi} \rangle \rightarrow 0$ and $\langle\sin{2\chi} \rangle \rightarrow 0 $, and the leakage would be
\begin{equation}
        \delta \hat{Q} \approx \frac{g_1}{2}\,I Q \,\langle \sin^2{2\chi} \rangle
            \langle \cos^2{2\chi} \rangle
\end{equation}
\begin{equation}
        \delta \hat{U} \approx \frac{g_1}{2}\,I U \,\langle \cos^2{2\chi} \rangle 
        \langle \cos^2{2\chi} \rangle 
\end{equation}
By evaluating pixel-averaged quantities in eq.~\eqref{eq:deltad2f}, we include the HWP synchronous signal from differential emission as well. Denoting now the perturbed signal as $d_{\mathrm{NL}}^{2f}$, eq.~\eqref{eq:pixav-nl-only} becomes
\begin{equation}
\label{eq:pixav-2f-nl}
    \begin{cases}
        \langle d_{NL}^{2f} \rangle = \langle d_{NL} \rangle + A_{2f} \mu 
        + g_1\big(A_{2f}^2 \nu + 2 I A_{2f} \mu + 2 Q A_{2f} \gamma + 2 U A_{2f} \eta\big), \\[4pt]
        \langle d_{NL}^{2f}\cos{2\chi} \rangle = \langle d_{NL}\cos{2\chi} \rangle + A_{2f} \gamma 
        + g_1\big(A_{2f}^2 \delta + 2 I A_{2f} \gamma + 2 Q A_{2f} \epsilon + 2 U A_{2f} \zeta\big), \\[4pt]
        \langle d_{NL}^{2f}\sin{2\chi} \rangle = \langle d_{NL}\sin{2\chi} \rangle + A_{2f} \eta 
        + g_1\big(A_{2f}^2 \kappa + 2 I A_{2f} \eta + 2 Q A_{2f} \zeta + 2 U A_{2f} \lambda\big),
    \end{cases}
\end{equation}
where we have introduced the averages
\begin{align*}
    \mu &= \langle\cos\Theta \rangle, & \nu &= \langle\cos^2\Theta \rangle, \\
    \gamma &= \langle\cos\Theta\cos2\chi \rangle, & \delta &= \langle\cos^2\Theta\cos2\chi \rangle, \\
    \epsilon &= \langle\cos\Theta\cos^2 2\chi \rangle, & \zeta &= \langle\cos\Theta\sin2\chi\cos2\chi \rangle, \\
    \eta &= \langle\cos\Theta\sin2\chi \rangle, & \kappa &= \langle\cos^2\Theta\sin2\chi \rangle, \\
    \lambda &= \langle\cos\Theta\sin^2 2\chi \rangle.
\end{align*}
and $\Theta = 2(\theta-\xi_{\mathrm{det}})$, $\xi_{\mathrm{det}}$ being the detector polarization angle, and $A_{2f}$ the amplitude of the synchronous HWP differential emission (in temperature units). The resulting total leakage maps are then:
\begin{equation}\label{eq:leakQ-total}
    \begin{split}
    \delta \hat{Q} = &-(B^2 - E) \Big( 
         g_1 \big( 
            A I^2 + 2 C I Q + 2 D I U + 2 \gamma I A_{2f} \\
        &\quad + F Q^2 + 2 H Q U + 2 \epsilon Q A_{2f} 
            + G U^2 + 2 \zeta U A_{2f} + \delta A_{2f}^2 
        \big) 
        + \gamma A_{2f} 
    \Big) \\
    &+ (A B - D) \Big( 
         \eta A_{2f} 
        + g_1 \big( 
            B I^2 + 2 D I Q + 2 E I U + 2 \eta I A_{2f} \\
        &\quad + H Q^2 + 2 G Q U + 2 \zeta Q A_{2f} 
            + J U^2 + 2 \lambda U A_{2f} + \kappa A_{2f}^2 
        \big) 
    \Big) \\
    &- (A E - B D) \Big( 
         g_1 \big( 
            I^2 + 2 A I Q + 2 B I U + 2 \mu I A_{2f} \\
        &\quad + C Q^2 + 2 D Q U + 2 \gamma Q A_{2f} 
            + E U^2 + 2 \eta U A_{2f} + \nu A_{2f}^2 
        \big) 
        + \mu A_{2f} 
    \Big)
    \end{split}
\end{equation}

\begin{equation}\label{eq:leakU-total}
    \begin{split}
    \delta \hat{U} = &- (A^2 - C) \Big( 
        \eta A_{2f} 
        + g_1 \big( 
            B I^2 + 2 D I Q + 2 E I U + 2 \eta I A_{2f} \\
        &\quad + H Q^2 + 2 G Q U + 2 \zeta Q A_{2f} 
            + J U^2 + 2 \lambda U A_{2f} + \kappa A_{2f}^2 
        \big) 
    \Big) \\
    &+ (A B - D) \Big( 
         g_1 \big( 
            A I^2 + 2 C I Q + 2 D I U + 2 \gamma I A_{2f} \\
        &\quad + F Q^2 + 2 H Q U + 2 \epsilon Q A_{2f} 
            + G U^2 + 2 \zeta U A_{2f} + \delta A_{2f}^2 
        \big) 
        + \gamma A_{2f} 
    \Big) \\
    &+ (A D - B C) \Big( 
        g_1 \big( 
            I^2 + 2 A I Q + 2 B I U + 2 \mu I A_{2f} \\
        &\quad + C Q^2 + 2 D Q U + 2 \gamma Q A_{2f} 
            + E U^2 + 2 \eta U A_{2f} + \nu A_{2f}^2 
        \big) 
        + \mu A_{2f} 
    \Big)
    \end{split}
\end{equation}

\begin{figure}[]
\centering
\includegraphics[width=\textwidth]{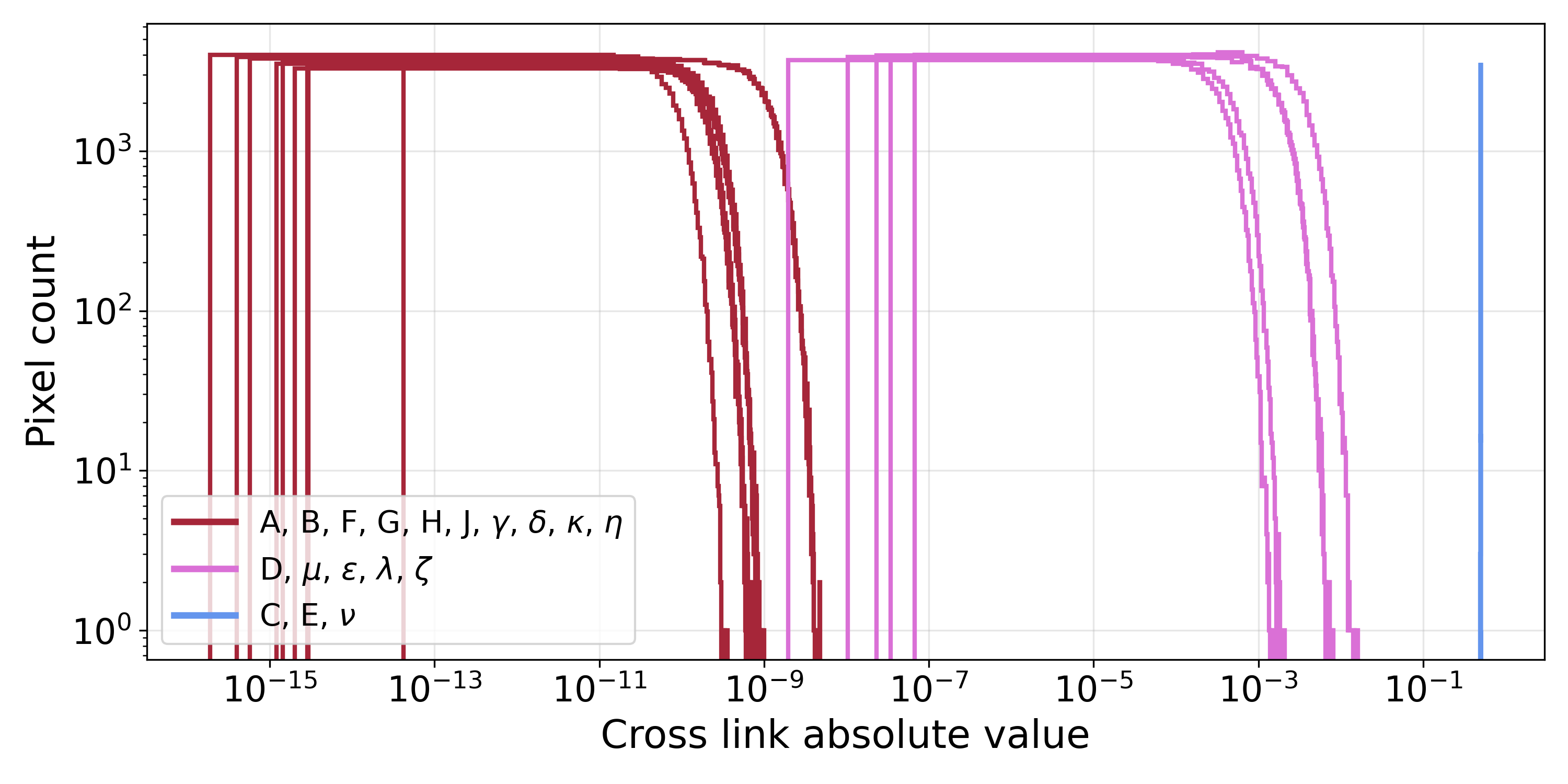}
\caption[Distribution of one-year pixel-averaged spin-weighted factors.]{Distribution of the pixel-averaged spin-weighted factors for a pair of orthogonal detectors observing the full sky for one year.}
\label{fig:allhists} 
\end{figure}

Once all terms in the map-making equation are written explicitly, many off-diagonal averages are seen to vanish (or be extremely small) for a well-conditioned cross-link matrix, i.e.\ when $\chi_j$ spans many values in $[0,2\pi]$ per pixel. Figure~\ref{fig:allhists} shows the distribution of cross-link matrix entries and the non-linear coupling coefficients for a pair of orthogonal detectors observing the full sky for one year. Figure~\ref{fig:allhists} highlights three regions in the absolute values of the cross-links: some terms are effectively negligible ($\mathcal{O}(10^{-10})$), while others are small but relevant ($\mathcal{O}(10^{-3})$). The latter contributes to the mixing of $Q$ and $U$ components. Given these scalings, we can speed up the production of contaminated maps by retaining only the leading terms. In this approximation, the final leakage maps become
\begin{equation}
    \begin{split}
        \delta \hat{Q} \approx & \; \frac{g_1}{\det M}\, E \big(2 I Q C + 2 A_{2f} Q \epsilon + 2 A_{2f} U \zeta\big) \\[4pt]
        = & \; \frac{g_1}{2}\,\langle \sin^2{2\chi} \rangle 
        \Big(
            I Q \langle \cos^2{2\chi} \rangle 
            + A_{2f} Q \langle \cos\Theta \cos^2{2\chi} \rangle 
            + A_{2f} U \langle \cos\Theta \sin{2\chi} \cos{2\chi} \rangle
        \Big),
    \end{split}
\end{equation}
\begin{equation}
    \begin{split}
        \delta \hat{U} \approx & \; \frac{g_1}{\det M}\, C \big(2 I U C + 2 A_{2f} Q \zeta + 2 A_{2f} U \lambda\big) \\[4pt]
        = & \; \frac{g_1}{2}\,\langle \cos^2{2\chi} \rangle 
        \Big(
            I U \langle \cos^2{2\chi} \rangle 
            + A_{2f} Q \langle \cos\Theta \sin{2\chi} \cos{2\chi} \rangle 
            + A_{2f} U \langle \cos\Theta \sin^2{2\chi} \rangle
        \Big).
    \end{split}
\end{equation}

\section{Validation: comparison of maps}
\label{sec:appendixmaps}

\begin{figure}[]
\centering
\includegraphics[width=\textwidth]{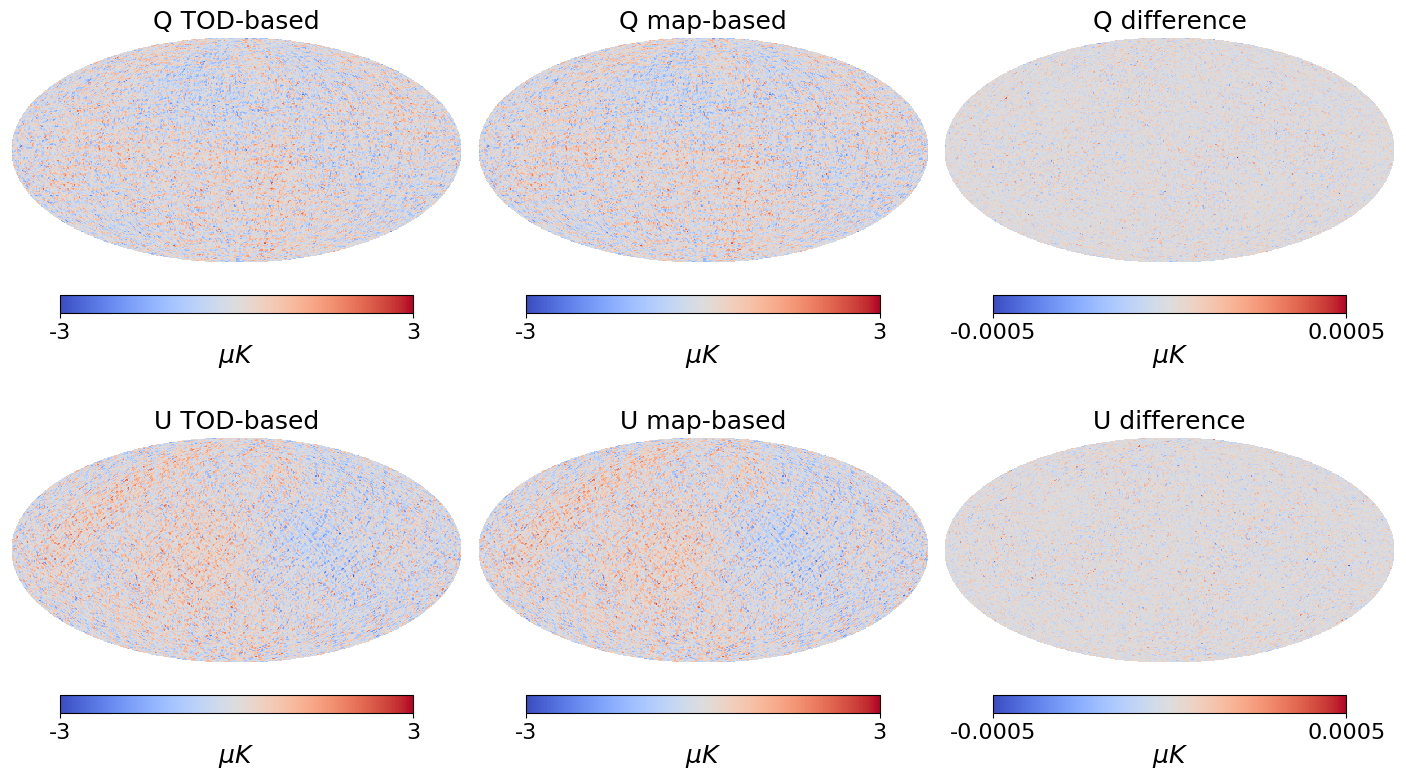}
\caption{Comparison between the $Q$ (upper) and $U$ (lower) polarization maps reconstructed from the TOD simulations and from the map-based framework. The third column shows the corresponding difference maps.}
\label{fig:maps} 
\end{figure}

We show in figure~\ref{fig:maps} the maps used to produce the spectra in the validation phase of this work, whose comparison is presented in figure~\ref{fig:spectra-comparison}. By employing the same scale, we can observe the high level of agreement between the two reconstruction methods, both for the Q and U polarization components. The difference maps highlight the residuals between the two approaches, which remain small and uniform across the sky, with no evident feature arising from the map-based procedure.

\acknowledgments
We acknowledge Guillaume Patanchon and Marta Monelli for useful discussion on the formalism presented this work. SM is supported by the European Union - Next Generation EU, Missione: I.4.1 Borse dottorati generici ricerca PNRR (Missione 4), Componente: 1, CUP 351: B83C22003210006. This work has also received funding by the European Union’s Horizon 2020 research and innovation program under grant agreement no. 101007633 CMB-Inflate. \textit{LiteBIRD} (phase A) activities are supported by the following funding sources: ISAS/JAXA, MEXT, JSPS, KEK (Japan); CSA (Canada); CNES, CNRS, CEA (France);
DFG (Germany); ASI (ASI Grants
No. 2020-9-HH.0 and 2016-24-H.1-2018), INFN, INAF (Italy); RCN (Norway); MCIN/AEI, CDTI (Spain); SNSA, SRC (Sweden); UKSA (UK); and NASA, DOE (USA). We acknowledge the use of computing resources at CINECA.

%
% =====================================================================================

% Bibliography

%% [A] Recommended: using JHEP.bst file
\bibliographystyle{JHEP}
\bibliography{biblio.bib}

\end{document}